\documentclass[aps,prx,10pt,twocolumn,superscriptaddress,longbibliography]{revtex4}
\usepackage{graphicx}
\usepackage{amssymb, amsmath}
\usepackage{color}
\usepackage{ulem}
\usepackage{bm}
\usepackage{graphicx}
\usepackage{subfigure}
\usepackage{dcolumn}
\usepackage{mathtools}
\usepackage{pifont}
\def\I{\uppercase\expandafter{\romannumeral 1}}
\def\II{\uppercase\expandafter{\romannumeral 2}}
\def\III{{\uppercase\expandafter{\romannumeral 3}}}
\def\IV{{\uppercase\expandafter{\romannumeral 4}}}
\def\V{{\uppercase\expandafter{\romannumeral 5}}}
\def\VI{{\uppercase\expandafter{\romannumeral 6}}}
\def\VII{{\uppercase\expandafter{\romannumeral 7}}}
\def\i{\lowercase\expandafter{\romannumeral 1}}
\def\ii{\lowercase\expandafter{\romannumeral 2}}
\def\iii{{\lowercase\expandafter{\romannumeral 3}}}
\def\iv{{\lowercase\expandafter{\romannumeral 4}}}
\def\v{{\lowercase\expandafter{\romannumeral 5}}}
\def\vi{{\lowercase\expandafter{\romannumeral 6}}}
\def\vii{{\lowercase\expandafter{\romannumeral 7}}}
\def\rr{\mathbf{r}}
\def\angstrom{\mbox{\normalfont\AA}}

\def\nn{\nonumber\\}

\def\angstrom{\mbox{\normalfont\AA}}
\def\k{\mathbf{k}}
\def\kt{\widetilde{\mathbf{k}}}
\def\q{\mathbf{q}}
\def\nn{\nonumber\\}

\def\s{\textrm{s}}

\newcommand{\xmark}{\ding{55}}
\begin{document}

\title{Theories for the correlated insulating states and quantum anomalous Hall phenomena in twisted bilayer graphene}

\author{Jianpeng Liu}
\affiliation{School of Physical Science and Technology, ShanghaiTech University, Shanghai 200031, China}
\affiliation{ShanghaiTech laboratory for topological physics, ShanghaiTech University, Shanghai 200031, China}
\affiliation{Department of Physics, Hong Kong University of Science and Technology, Kowloon, Hong Kong, China}

\author{Xi Dai}
\affiliation{Department of Physics, Hong Kong University of Science and Technology, Kowloon, Hong Kong, China}

\begin{abstract}
The experimentally observed correlated insulating states and quantum anomalous Hall (QAH)  effect in twisted bilayer graphene (TBG) have drawn significant attention. However, up to date, the specific mechanisms of these intriguing phenomena are still open questions. Using a fully unrestricted Hartree-Fock variational method, we have successfully explained the correlated insulating states and QAH effects at all integer fillings of the flat bands in TBG. 
Our results indicate that states breaking flavor (valley and spin) symmetries are energetically favored at all integer fillings.
In particular,
the correlated insulating states at $\pm 1/2$ filling and at the charge neutrality point are all valley polarized sates which break $C_{2z}$ and time-reversal ($\mathcal{T}$) symmetries, but preserves  $C_{2z}\mathcal{T}$ symmetry. 
Such valley polarized states exhibit  ``moir\'e orbital antiferromagnetic ordering" on an emergent honeycomb lattice  with compensating circulating current pattern in the moir\'e supercell. Such moir\'e orbital antiferromagnetic states are unprecedented in condensed matter physics.   
Within the same theoretical framework, our calculations indicate that the $C\!=\!\mp 1$ QAH states at $\pm 3/4$ fillings of the magic-angle TBG are  spin and orbital ferromagnetic states, which emerge when a staggered sublattice potential is present. We find that the nonlocalness of the exchange interactions tend to enhance the bandwidth of the low-energy bands due to the exchange-hole effect, which reduces the gaps of the correlated insulator phases. The nonlocal exchange interactions also dramatically enhance the spin polarization of the system, which significantly stabilize the orbital and spin ferromagnetic QAH state at $3/4$ filling of TBG aligned with hexagonal boron nitride (hBN). We also predict that, by virtue of the orbital ferromagnetic nature, the QAH effects at electron and hole fillings of hBN-aligned TBG would exhibit hysteresis loops with opposite chiralities.
\end{abstract}

\maketitle

The moir\'e graphene system has been an exciting area of condensed matter physics since the discoveries of the correlated insulating states \cite{cao-nature18-mott,efetov-nature19,tbg-stm-pasupathy19,tbg-stm-andrei19,tbg-stm-yazdani19, tbg-stm-caltech19, young-tbg-science19} and unconventional superconductivity \cite{cao-nature18-supercond,dean-tbg-science19,marc-tbg-19, efetov-nature19} in twisted bilayer graphene (TBG). At small twist angles, for each spin the low-energy states of TBG can be characterized by four bands contributed by the two nearly decoupled valleys $K$ and $K'$ \cite{santos-tbg-prl07,macdonald-pnas11,po-prx18, yuan-prb18, koshino-prx18, kang-prx18}. Around the first magic angle $\sim 1.05^{\circ}$, the bandwidths of the four low-energy bands become vanishingly small, and these nearly flat bands are believed to be responsible for
the intriguing correlation effects observed in TBG. These flat bands in magic-
angle TBG were also found to exhibit nontrivial topological properties \cite{song-tbg-prl19, po-tbg-prb19, yang-tbg-prx19, jpliu-prb19} . In particular, it has been proposed by the authors that the two flat bands (per spin per valley) are just the two zeroth pseudo Landau levels (LLs) resulting from some opposite pseudo magnetic fields generated by the moir\'e potential, which naturally carry nonzero Chern numbers $\pm 1$ \cite{jpliu-prb19}.  Recently, different types of  insulating states have been found at different integer fillings\cite{efetov-nature19}, and some of these correlated insulating states even exhibit QAH effects with Chern number $\pm 1$ (3/4 filling) \cite{sharpe-science-19, young-tbg-science19} and $\pm 2$ ($\pm$ 1/2 filling) \cite{efetov-tbg-chern}.
Although various theoretical models have been proposed already to explain the different insulating states at each particular filling \cite{po-prx18, isobe-prx18, xu-lee-prb18, huang-tbg-sb19,liu-prl18,rademaker-prb18, venderbos-prb18, kang-tbg-prl19, xie-tbg-2018,zaletel-tbg-2019, zhang-senthil-tbg19,wu-tbg-collective, chatterjee-tbg-arxiv2019, repellin-tbg-prl20,zaletel-tbg-hf}, up to date there is still no unified theory which can explain all the insulating and QAH phenomena at different integer fillings. 

On the other hand,  for a TBG system, the most general model to describe its electronic structure is a 2D interacting Dirac Fermi gas with spin, valley, layer and sublattice degrees of freedom under the modulation of moir\'e potential. Similar to the 3D inhomogeneous electron-gas problem being considered in density functional theory (DFT) \cite{dft1} for ordinary crystals, such a general model for TBG can be treated by the DFT on the moir\'e length scale.
In the present study, we simply take the exchange-correlation energy functional as the Hartree-Fock energy and our approach is reduced to the unrestricted Hartree-Fock variational method.  Such an approach can be further improved if the Hartree-Fock energy is replaced by more accurate functional extracted from quantum Monte-Carlo simulation.

Using such a fully unrestricted Hartree-Fock (HF) variational method applied to all energy bands of TBG,  we have successfully explained the correlated insulating states at $\pm1/2$ and $0$ fillings \cite{cao-nature18-mott,tbg-stm-andrei19,tbg-stm-caltech19,efetov-nature19}, as well as the QAH effect at $3/4$ filling \cite{young-tbg-science19, sharpe-science-19}  when the hBN substrate is aligned with TBG. Our results indicate that  states breaking the flavor symmetry, i.e., the valley and/or spin polarized states, are energetically favored at all integer fillings, and would lead to diverse phenomena at different fillings. In particular, the valley polarized correlated insulating states at $\pm 1/2$ filling and at the charge neutrality point  exhibit opposite circulating current loops, which generate staggered orbital magnetic fluxes on an emergent honeycomb lattice in the moir\'e supercell. To the best of our knowledge, such  ``moir\'e  orbital antiferromagnetic" states on an emergent honeycomb lattice have never been proposed nor discussed in literatures. These states break both orbital time-reversal ($\mathcal{T}$) and $C_{2z}$ symmetries which exhibit giant nonlinear optical responses induced by the current-loop order. On the other hand, such states still exhibit vanishing anomalous Hall effect due to the combined $C_{2z}\mathcal{T}$ symmetry. 

When an hBN substrate is aligned with the TBG, the HF ground states with local-exchange approximation at $\pm 1/2$ filling are valley polarized and spin degenerate states exhibiting QAH effects with Chern number ($C$) $\mp 2$. 
The $C=\mp 2$ QAH states can be stabilized by vertical magnetic fields due to the orbital magnetic Zeeman effect, but would be suppressed by in-plane magnetic fields due to (interaction-enhanced) spin Zeeman effect. On the other hands, the nonlocal exchange interactions tend to enhance the spin polarization such that the ground states at $\pm 1/2$ fillings of hBN-aligned TBG become spin polarized insulators with zero Chern number.

Within the same theoretical framework, we have also studied the QAH states at $\pm3/4$ filling of the flat bands in hBN-aligned TBG at the magic angle. Our results indicate that the $C\!=\!\mp 1$ QAH state at $\pm 3/4$ filling is  a state with co-existing spin and orbital ferromagnetic orders, which can be significantly stabilized by a nonzero staggered sublattice potential, and also by nonlocal exchange interactions due to the enhanced spin polarization. Moreover, for the same valley polarization, the Chern numbers at the electron and hole fillings are opposite in sign, but the orbital magnetizations have the same sign,   which implies that the hysteresis loops of the QAH states at electron and hole fillings would have  opposite chiralities.

\section{Non-interacting bandstructures}
\begin{figure}
\includegraphics[width=3.5in]{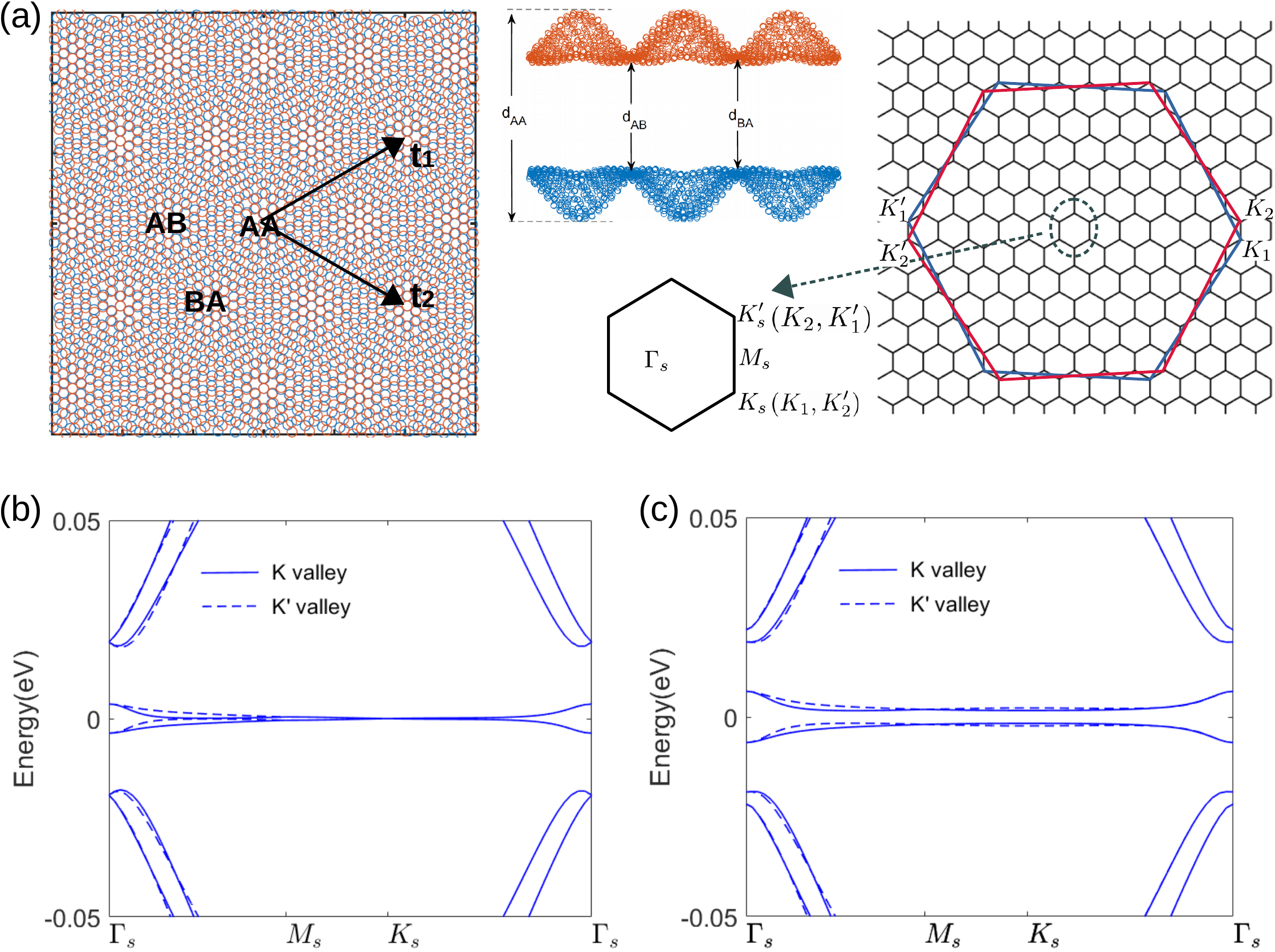}
\caption{(a) The lattice structure and Brillouin zone of the twisted bilayer graphene (TBG). The bandstructure of TBG at the magic angle $\theta=1.05^{\circ}$ are shown in (b)-(c): (b) without the hBN substrate, and (c) aligned with the hBN substrate. See text for more details.}
\label{fig:lattice}
\end{figure}

Moir\'e pattern is formed when the two graphene monolayers are twisted with respect to each other by some commensurate angle $\theta$.
The moir\'e lattice constant $L_s\!=\!a/(2\sin{\theta/2})$ is much greater than the monolayer graphene lattice constant $a$ at small twist angles, as shown in Fig.~\ref{fig:lattice}. The low-energy states of TBG at small twist angles can be  well described by the continuum model proposed in Ref.~\onlinecite{macdonald-pnas11}.  
The states around either  $K$ or $K'$ valleys of graphene can  be separately folded into the moir\'e BZ, leading to valley degeneracy in additional to the spin degeneracy.  Moreover, at small twist angles,
the coupling between the low-energy states around the two valleys can be neglected at small twist angles \cite{macdonald-pnas11, castro-neto-prb12}, and the  charge is separately conserved for each valley, leading to an emergent valley $U(1)$ symmetry (dubbed as $U_v(1)$ symmetry). Moreover, as spin-orbit coupling is negligible in graphene, there is separate spin $SU(2)$ symmetry for each valley, such that the entire system has a $U_v(1)\times SU(2)\times SU(2)$ symmetry.  In addition to these continuous symmetries, for each valley the continuum Hamiltonian (see Eq.~(\ref{eq:h0-mu})  has $C_{3z}$, $C_{2z}\mathcal{T}$,  and $C_{2x}$ symmetries, where $\mathcal{T}$ is the time-reversal operation for spinless fermions (i.e., complex conjugation). The two valleys can be transformed to each other by  $\mathcal{T}$, $C_{2z}$, and $C_{2y}$ operations.

The non-interacting bandstructures of TBG  at the magic angle are shown in Fig.~\ref{fig:lattice}(b), where the solid and dashed lines represent those of the $K$ and $K'$ valleys respectively. The Fermi velocities at the magic angle are vanishing, and the overall bandwidth is less than $10\,$meV. If the hBN substrate is aligned with TBG,  a staggered sublattice potential is exerted on the bottom layer graphene, which  breaks $C_{2z}\mathcal{T}$ symmetry and opens a gap $\sim 4\,$meV at the $K_s$ and $K_s'$ points as shown in Fig.~\ref{fig:lattice}(c). As a result, the valence and conduction band flat bands acquire  nonzero valley Chern numbers $\pm 1$.

\section{Coulomb interactions and the spontaneous symmetry breaking}
In order to understand the experimentally observed correlated insulating states, we apply screened Coulomb interactions to all energy bands of TBG. The Coulomb interactions between the electrons in TBG can be expressed in momentum space as
\begin{align}
H_{C}=\frac{1}{2N_s}\sum_{\alpha\alpha'}\sum_{\k\k'\q}\sum_{\sigma\sigma'} V(\q)\, \hat{c}_{\k+\q, \alpha\sigma}^{\dagger}\hat{c}_{\k'-\q, \alpha'\sigma'}^{\dagger} \hat{c}_{\k',\alpha'\sigma'} \hat{c}_{\k,\alpha\sigma}
\label{eq:coulomb}
\end{align}
where $\k$ is the atomic wavevector expanded around the Dirac point in the Brillouin zone of monolayer graphene, which can be written as $\k=\widetilde{\k}+\mathbf{G}$, where $\widetilde{\k}$ denotes the wavevector with the moir\'e Brillouin zone and $\mathbf{G}$ is a moir\'e reciprocal lattice vector.
 $\alpha$ is the combined layer and sublattice index, and $\sigma$ is the spin index.  $c^{\dagger}_{\k,\alpha\sigma}$ and $c_{\k,\alpha\sigma}$ represent the creation and annihilation operators of the Dirac fermions. $N_s$ is the total number of moir\'e supercells in the entire system. $V(\q)$ is Fourier transform of the Coulomb interaction
\begin{align}
V(\q)= \frac{1}{\Omega_{M}}\int d\rr\,\frac{e^2\,e^{-\kappa\vert\rr\vert}}{4\pi\epsilon\epsilon_0 \vert\mathbf{\rr}\vert}e^{-i\q\cdot\rr}\;,
\label{eq:vq}
\end{align}
where $\Omega_M$ is the area of the moir\'e supercell, $\kappa$ is introduced as the inverse screening length, and $\epsilon$ is the  background dielectric constant. $\kappa$ and $\epsilon$ will be treated as two free parameters in the following calculations. We will construct phase diagrams in the parameter space spanned by $(\kappa, \epsilon)$.

We apply the Hartree-Fock variational method, in which the following mean field Hamiltonian, 
\begin{align}
H_{\textrm{MF}}=&H_0 + \sum_{\k,\mathbf{Q},\gamma} 
O^H({\mathbf{Q}})\,c^{\dagger}_{\k + \mathbf{Q},\gamma}\;c_{\k,\gamma}\;\nn
&+ \sum_{\k,\gamma,\gamma'} O^F_{\gamma \gamma'} (\k)\;c^{\dagger}_{\k,\gamma} \,c_{\k,\gamma'}
\label{eq:hmf}
\end{align}
has been proposed to determine the variational ground state wave function $\vert\psi\{O^{HF}\}\rangle$ (with $O^{HF}\!\equiv\!\{O^{H}(\mathbf{Q}),O^{F}_{\gamma\gamma'}\}$), where $H_0$ is the non-interacting Hamiltonian for TBG, and the matrices $O^H(\mathbf{Q})$ and $O^F_{\gamma \gamma'}(\k)$  denote the Hartree and Fock types of terms respectively with $\mathbf{Q}$ being the moir\'e reciprocal lattice vectors, and the indices $\gamma$  and  $\gamma'$ denoting the combined indices of valley, spin and sublattice degrees of freedom. These Hartree-Fock order parameters will be determined by minimizing the total  energy functional $E_{T}=\langle\psi\{O^{HF}\}\vert H_0+H_C\vert\psi\{O^{HF}\}\rangle$, where $H_C$ denotes electrons' Coulomb interactions in TBG.
More details about the Hartree-Fock variational method are given in Appendix~\ref{sec:append3}.  We make the approximation that the system preserves moir\'e superlattice translational symmetry, which is supported by recent scanning tunneling microscopy (STM) measurements \cite{tbg-stm-caltech19,tbg-stm-pasupathy19, tbg-stm-yazdani19, tbg-stm-andrei19}. 
%We make two approximations:  (\i) the system preserves moir\'e superlattice translational symmetry, which is supported by recent scanning tunneling microscopy (STM) measurements \cite{tbg-stm-caltech19,tbg-stm-pasupathy19, tbg-stm-yazdani19, tbg-stm-andrei19}. (\ii) We do not consider the spontaneous breaking of $C_{3z}$ symmetries. This is because the TBG system typically has heterostrain, which naturally break the $C_{3z}$ symmetry at the single-particle level. Thus it is unclear the nematicity observed in Refs.~\onlinecite{tbg-stm-andrei19, tbg-stm-pasupathy19} is driven purely by Coulomb interactions, or driven by heterostrain (and may be enhanced by Coulomb interactions). Previous theoretical calculations show that nematicity shows up only when $C_{3z}$ symmetry is already broken in the non-interacting Hamiltonian \cite{ashvin-nematic-arxiv19}. On the other hands,  nematicity can naturally co-exist with other order parameters induced by the breaking of valley, spin, and/or sublattice symmetries, and the latter are responsible for the correlated insulator phase and quantum anomalous Hall effect according to our calculations.

With the moir\'e translational symmetry,  all the HF order parameters can be divided into two categories : the valley polarized (VP) order and the intervalley coherent (IVC) order. The VP order can be generically written as $\tau_z s_i \sigma_j$ and $\tau_0 s_i \sigma_j$ ($i, j = 0, x, y, z$),  where $\mathbf{\tau}$, $\mathbf{s}$ and $\mathbf{\sigma}$ are the Pauli matrices representing the valley, spin, and sublattice degrees of freedom of the system.
The VP-type order parameters preserve the valley charge conservation, but may break global spin rotational symmetry, $\mathcal{T}$, $C_{2z}$ and/or $C_{2z}\mathcal{T}$ symmetry. 
On the other hand, the IVC order always breaks the valley  charge-conservation ($U_v(1)$ symmetry), which has the form $\tau_{x} s_i \sigma_j$ and/or $\tau_y s_i \sigma_j$ ($i, j=0, x, y, z$). The IVC order may further break the relative spin rotational symmetry between the two valleys dubbed as $SU_v(2)$ and even the global $SU(2)$ symmetry (dubbed as $SU_g(2)$).  Since the atomic spin-orbit coupling in carbon is negligible, the $\mathcal{T}$ symmetry here refers to the orbital time-reversal symmetry. The spin degrees of freedom is not involved in the time-reversal operation. On the other hand, since we assume the Fock order parameters are independent of the layer degrees of freedom, we do not specifically discuss the spontaneous breaking of $C_{2x}$ and $C_{2y}$ symmetries. One should keep in mind that order parameters which induce staggered sublattice potential, such as $\tau_i s_j \sigma_z$, also break $C_{2x}$ symmetry, since the $A$ and $B$ sublattices of opposite layers are connected to each other by $C_{2x}$ operation; while order parameters which induce valley polarizations, such as $\tau_z s_i \sigma_j$, also break $C_{2y}$ symmetry, since the two valleys are  transformed to each other by $C_{2y}$ operation. It is worthwhile to note that states breaking $C_{3z}$ symmetry have also been taken into account, since some of the VP and/or IVC order parameters such as $\tau_{0,z} s_0\sigma_{x,y}$ and $\tau_{x,y} s_0 \sigma_{0,z}$ already break $C_{3z}$ symmetry. However, our calculations indicate the $C_{3z}$ symmetry breaking is not necessary in opening a gap at any integer filling.  In other words, whether $C_{3z}$ is broken or not does not change the nature of the correlated insulators at 0 and 1/2 fillings. Moire details are to be discussed in Sec.~\ref{sec:local-exchange} and Sec.~\ref{sec:nonlocal-exchange}.
 
%It is also worthwhile to note that some of the VP and/or IVC orders (such as $\tau_z s_0\sigma_x$, $\tau_x s_0\sigma_z$) can also break $C_{3z}$ symmetry. With local exchange approximation (Sec.~\ref{sec:local-exchange}), the $C_{3z}$-breaking effects are manifested only on the atomic length scale due to the locality of the exchange interactions, the low-energy states still (approximately) exhibit $C_{3z}$ symmetry on the moir\'e length scale. Including the nonlocalness of the exchange interactions would induce finite correlation length (on the order of the moir\'e lattice constant), thus may break $C_{3z}$ symmetry on the moir\'e length scale. However, our calculations indicate the $C_{3z}$ symmetry breaking is not necessary in opening a gap at any integer filling.  In other words, whether $C_{3z}$ is broken or not does not change the nature of the correlated insulators at 0 and 1/2 fillings. Moire details are to be discussed in Sec.~\ref{sec:local-exchange} and Sec.~\ref{sec:nonlocal-exchange}.

\begin{figure}
\includegraphics[width=3.0in]{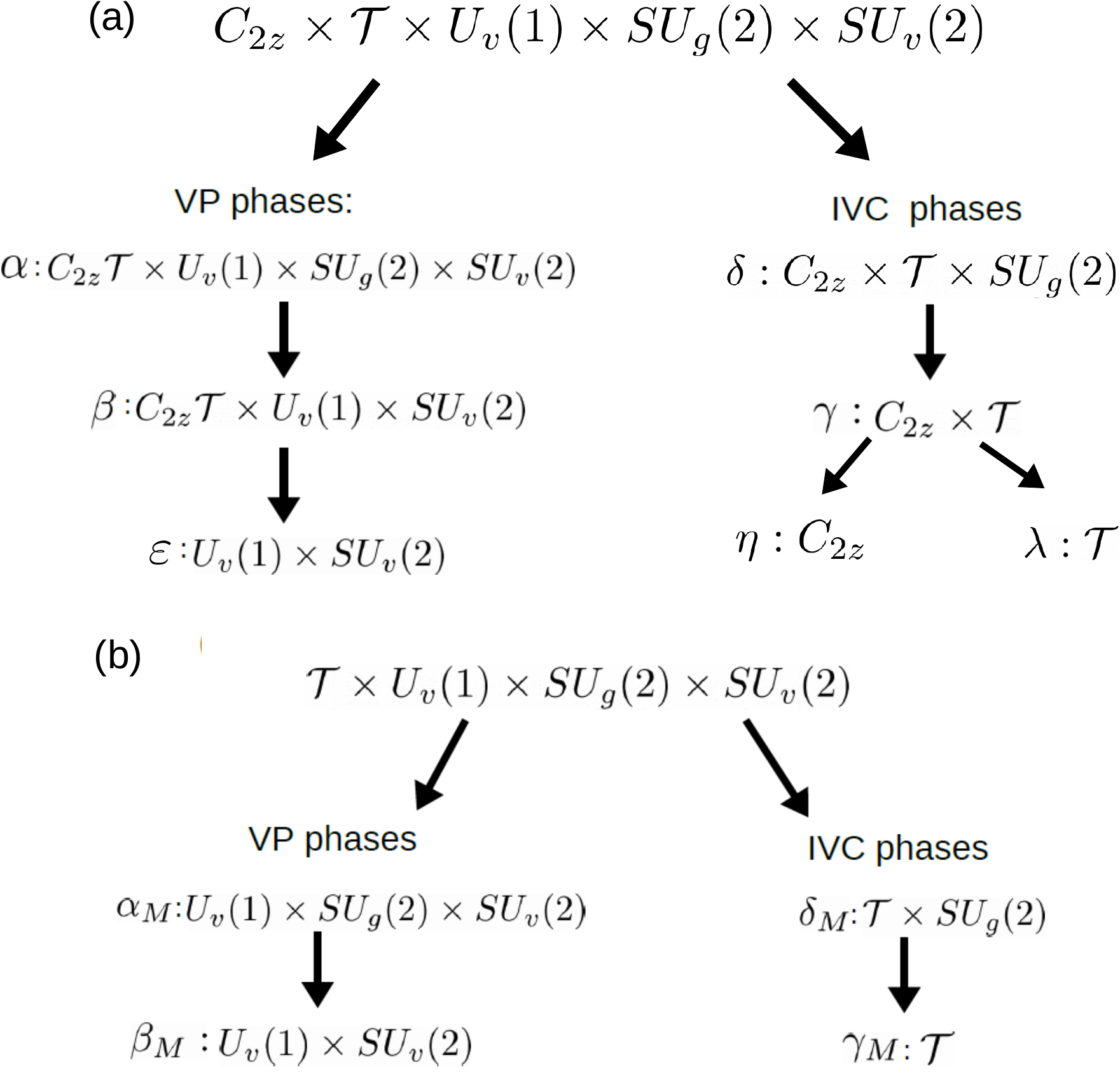}
\caption{The two paths of spontaneous symmetry breakings in twist bilayer graphene: (a) without the alignment of hBN substrate, and (b) with the alignment of hBN substrate. See text for more discussions.} 
\label{fig:symmbreak}
\end{figure}

There are two different paths of spontaneous symmetry breakings in  the TBG system: one is through the generation of the VP-type orders, and the other is through the IVC-type orders, which are schematically shown in Fig.~\ref{fig:symmbreak}(a). The full symmetry groups of TBG without hBN alignment is $C_{2z}\times\mathcal{T}\times U_v(1)\times SU_g(2)\times SU_v(2)$. As already explained above, we do not discuss the spontaneous breaking of $C_{3z}$, $C_{2x}$ and $C_{2y}$ symmetries. 
For VP-type $\alpha$ phase, the $C_{2y}$, $C_{2z}$, and $\mathcal{T}$ are broken, but $C_{2z}\mathcal{T}$ as a combined symmetry is still preserved, so do other continuous symmetries. 
The $SU_g(2)$ symmetry can be further broken based on the $\alpha$ phase, which becomes the $\beta$ phase. If the $C_{2z}\mathcal{T}$ symmetry is further broken in the $\beta$ phase, the system becomes the $\epsilon$ phase as shown in Fig.~\ref{fig:symmbreak}(a). On the other hand, there is another path of spontaneous symmetry breaking (SSB) through the IVC-type orders. The system can first break $U_{v}(1)\times SU_v(2)$ symmetry, with three Goldstone modes from the valley charge and spin fluctuations, and such a phase is dubbed as ``$\delta$" phase in Fig.~\ref{fig:symmbreak}(a). The $SU_g(2)$ symmetry can be further broken in the $\delta$ phase, which becomes the ``$\gamma$ phase", in which there are five Goldstone modes from the global spin fluctuations, valley spin fluctuations, and valley charge fluctuations. Either $C_{2z}$ or $\mathcal{T}$ symmetry can be further broken starting from the $\gamma$ phase, leading to the $\lambda$ or $\eta$ phase as shown in Fig.~\ref{fig:symmbreak}(a).
Similar paths can be drawn for TBG aligned with hBN substrate as shown in Fig.~\ref{fig:symmbreak}(b). The difference is that for hBN-aligned TBG $C_{2z}$ (thus $C_{2z}\mathcal{T}$) is already broken in the non-interacting Hamiltonian. The corresponding SSB phases are denoted by $\alpha_{M}$, $\beta_M$, $\delta_M$, and $\gamma_M$ in Fig.~\ref{fig:symmbreak}(b). The symmetry-allowed Fock order parameters in the valley-spin-sublattice space in all the different phases are enumerated in Table~\ref{table:op-all}.

\begin{table}[bth]
\caption{Different spontaneous-symmetry-breaking phases of the TBG system. $N_{\textrm{gst}}$ denotes the number of Goldstone modes.}
\begin{ruledtabular}
\begin{tabular}{lclclclc}
 & $\mathcal{T}$ & $C_{2z}$  &  $C_{2z}\mathcal{T}$   &  $U_v(1)$ & $SU_g(2)$ & $SU_v(2)$ & $N_{\textrm{gst}}$ \\
$\alpha$  & \xmark & \xmark & $\checkmark$   & $\checkmark$  &  $\checkmark$  & $\checkmark$   & 0  \\ 
$\beta$  & \xmark  & \xmark  & $\checkmark$   & $\checkmark$  &  \xmark  & $\checkmark$   & 2  \\
$\gamma$  & $\checkmark$  & $\checkmark$  & $\checkmark$ & \xmark  &  \xmark  & \xmark   & 5  \\
$\delta$  & $\checkmark$  & $\checkmark$ &  $\checkmark$   & \xmark  &  $\checkmark$  & \xmark   & 3  \\
$\eta$  & \xmark  & $\checkmark$  & \xmark & \xmark  &  \xmark  & \xmark   & 5  \\
$\lambda$  & $\checkmark$  & \xmark &  \xmark   & \xmark  &  \xmark  & \xmark   & 5  \\
$\varepsilon$  & \xmark  & \xmark  & \xmark   & $\checkmark$  &  \xmark  & $\checkmark$   & 2  \\
$\alpha_M$  & \xmark & - & -   & $\checkmark$  &  $\checkmark$  & $\checkmark$   & 0  \\
$\beta_M$  & \xmark  & - & -  & $\checkmark$   &  \xmark  & $\checkmark$   & 2  \\
$\gamma_M$  & $\checkmark$ & - & -   & \xmark  &  \xmark  & \xmark   & 5  \\
$\delta_M$  & $\checkmark$ & -  & -   & \xmark  &  $\checkmark$  & \xmark  & 3 
\end{tabular}
\end{ruledtabular}
\label{table:order-parameter}
\end{table}
\begin{figure*}
\includegraphics[width=5.1in]{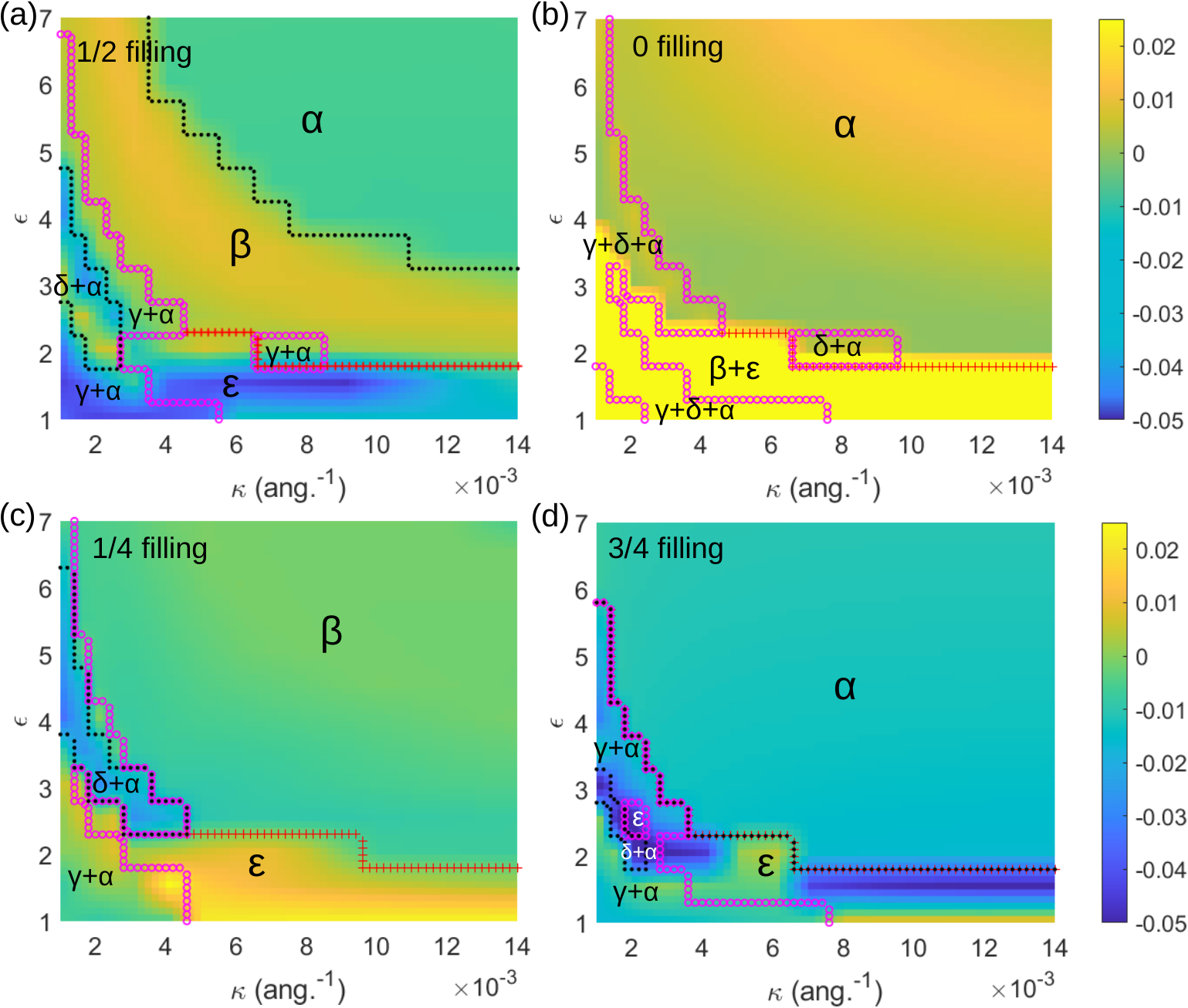}
\caption{The indirect gaps (in units of eV) of the Hartree-Fock ground states (with local-exchange approximation) of magic-angle TBG (without hBN alignment) at different fillings: (a) at 1/2 filling,  (b) at 0 filling, (c) at 1/4 filling, (d) at 3/4 filling.} 
\label{fig:phasediagram}
\end{figure*}

\begin{table}[bth]
\caption{Symmetry-allowed order parameters for different phases}
\begin{ruledtabular}
\begin{tabular}{lc}
$\alpha$ & $\tau_0 s_0\sigma_0$, $\tau_0 s_0\sigma_x$, $\tau_0 s_0\sigma_y$, $\tau_z s_0\sigma_0$, $\tau_z s_0\sigma_x$, $\tau_z s_0\sigma_y$ \\
\hline
$\beta$  & $\tau_0 s_0\sigma_0$, $\tau_0 s_0\sigma_x$, $\tau_0 s_0\sigma_y$, $\tau_z s_0\sigma_0$, $\tau_z s_0\sigma_x$, $\tau_z s_0\sigma_y$ \footnote{$\mathbf{s}_{\hat{\mathbf{n}}}$ represents spin operator pointing along an arbitrary axis $\hat{\mathbf{n}}$.} \\
&  $\tau_0 \mathbf{s}_{\hat{\mathbf{n}}} \sigma_0$, $\tau_0 \mathbf{s}_{\hat{\mathbf{n}}} \sigma_x$, $\tau_0 \mathbf{s}_{\hat{\mathbf{n}}} \sigma_y$,
$\tau_z \mathbf{s}_{\hat{\mathbf{n}}} \sigma_0$, $\tau_z \mathbf{s}_{\hat{\mathbf{n}}} \sigma_x$, $\tau_z \mathbf{s}_{\hat{\mathbf{n}}} \sigma_y$\\
\hline
$\varepsilon$ &   $\tau_0 s_0\sigma_j$, $\tau_0 \mathbf{s}_{\hat{\mathbf{n}}}\sigma_j$, $\tau_z s_0\sigma_j$, $\tau_z \mathbf{s}_{\hat{\mathbf{n}}}\sigma_j$, ($j=0, x, y, z$) \\ 
\hline
$\delta, \gamma$ & $\tau_x s_0 \sigma_0$, $\tau_x \mathbf{s}_{\hat{\mathbf{n}}}\sigma_0$,
$\tau_x s_0 \sigma_x$, $\tau_x \mathbf{s}_{\hat{\mathbf{n}}}\sigma_x$, $\tau_y s_{0}\sigma_z$, $\tau_y \mathbf{s}_{\hat{\mathbf{n}}}\sigma_z$  \footnote{The $\tau_{x/y} s_0$-type and $\tau_{x/y}\mathbf{s}_{\hat{\mathbf{n}}}$-type IVC order parameters can co-exist in the $\gamma$ ($\gamma_M$) phase, but they cannot coexist in the $\delta$ ($\delta_M$) phase.}  \\
\hline
$\eta$ & $\tau_x s_0 \sigma_0$, $\tau_x \mathbf{s}_{\hat{\mathbf{n}}}\sigma_0$,
$\tau_x s_0 \sigma_x$, $\tau_x \mathbf{s}_{\hat{\mathbf{n}}}\sigma_x$, \\
& $\tau_y s_{0}\sigma_y$, $\tau_y \mathbf{s}_{\hat{\mathbf{n}}}\sigma_y$,  $\tau_y s_{0}\sigma_z$, $\tau_y \mathbf{s}_{\hat{\mathbf{n}}}\sigma_z$ \\
\hline
$\lambda$ & $\tau_x s_0\sigma_0$, $\tau_x s_0\sigma_x$, $\tau_x s_0\sigma_z$,
$\tau_y s_0\sigma_0$, $\tau_y s_0\sigma_x$, $\tau_y s_0\sigma_z$ \\
  &  $\tau_x \mathbf{s}_{\hat{\mathbf{n}}}\sigma_0$, $\tau_x \mathbf{s}_{\hat{\mathbf{n}}}\sigma_x$, $\tau_x \mathbf{s}_{\hat{\mathbf{n}}}\sigma_z$,
$\tau_y \mathbf{s}_{\hat{\mathbf{n}}}\sigma_0$, $\tau_y \mathbf{s}_{\hat{\mathbf{n}}}\sigma_x$, $\tau_y \mathbf{s}_{\hat{\mathbf{n}}}\sigma_z$\\
\hline
$\alpha_M$ & $\tau_0 s_0\sigma_j$, $\tau_z s_0\sigma_j$, ($j=0,x,y,z$) \\
\hline
$\beta_M$ & $\tau_0 s_0\sigma_j$, $\tau_0 \mathbf{s}_{\hat{\mathbf{n}}}\sigma_j$, $\tau_z s_0\sigma_j$, $\tau_z \mathbf{s}_{\hat{\mathbf{n}}}\sigma_j$, ($j=0, x, y, z$) \\
\hline
$\delta_M, \gamma_M$  & $\tau_x s_0\sigma_0$, $\tau_x s_0\sigma_x$, $\tau_x s_0\sigma_z$,
$\tau_y s_0\sigma_0$, $\tau_y s_0\sigma_x$, $\tau_y s_0\sigma_z$ \\
  &  $\tau_x \mathbf{s}_{\hat{\mathbf{n}}}\sigma_0$, $\tau_x \mathbf{s}_{\hat{\mathbf{n}}}\sigma_x$, $\tau_x \mathbf{s}_{\hat{\mathbf{n}}}\sigma_z$,
$\tau_y \mathbf{s}_{\hat{\mathbf{n}}}\sigma_0$, $\tau_y \mathbf{s}_{\hat{\mathbf{n}}}\sigma_x$, $\tau_y \mathbf{s}_{\hat{\mathbf{n}}}\sigma_z$
\end{tabular}
\end{ruledtabular}
\label{table:op-all}
\end{table}

\section{Phase diagrams with local exchange approximations}
\label{sec:local-exchange}

We first consider the limit of local exchange interactions. The real-space locality implies that the Fock order parameters $O^{F}_{\gamma\gamma'}(\k)=\int d\mathbf{r}\, e^{-i\mathbf{k}\cdot\mathbf{r}}\,O^{F}_{\gamma\gamma'}(\mathbf{r})\delta(\mathbf{r})=O^{F}_{\gamma\gamma'}(\mathbf{0})$ are independent of the wavevector $\k$, and are only dependent on the valley, spin and sublattice degrees of freedom denoted by indices $\gamma$ and $\gamma'$. The interaction-driven spontaneous symmetry breaking in the valley-spin-sublattice space can be well captured within the local-exchange approximation. Including the nonlocalness of the Fock term would induce the ``exchange-hole" effects, i.e., two electrons with the same flavor degrees of freedom tend to avoid each other in real space due to Pauli exclusion principle, which effectively amplify the kinetic energy of electrons thus enhances the bandwidth. Moreover, in TBG system it turns out that the nonlocalness of the exchange interactions tend to increase the spin splittings of the flat bands, which significantly stabilizes the QAH states at 3/4 filling for hBN-aligned TBG.
In this section we first discuss the correlated insulating states and quantum anomalous Hall phenomena with such local-exchange approximation. Effects of nonlocal exchange will be discussed in Sec.~\ref{sec:nonlocal-exchange}.

\subsection{Correlated insulators at zero and half fillings in TBG without hBN alignment}
\label{sec:local-exchange-a}

\begin{figure}
\includegraphics[width=2.8in]{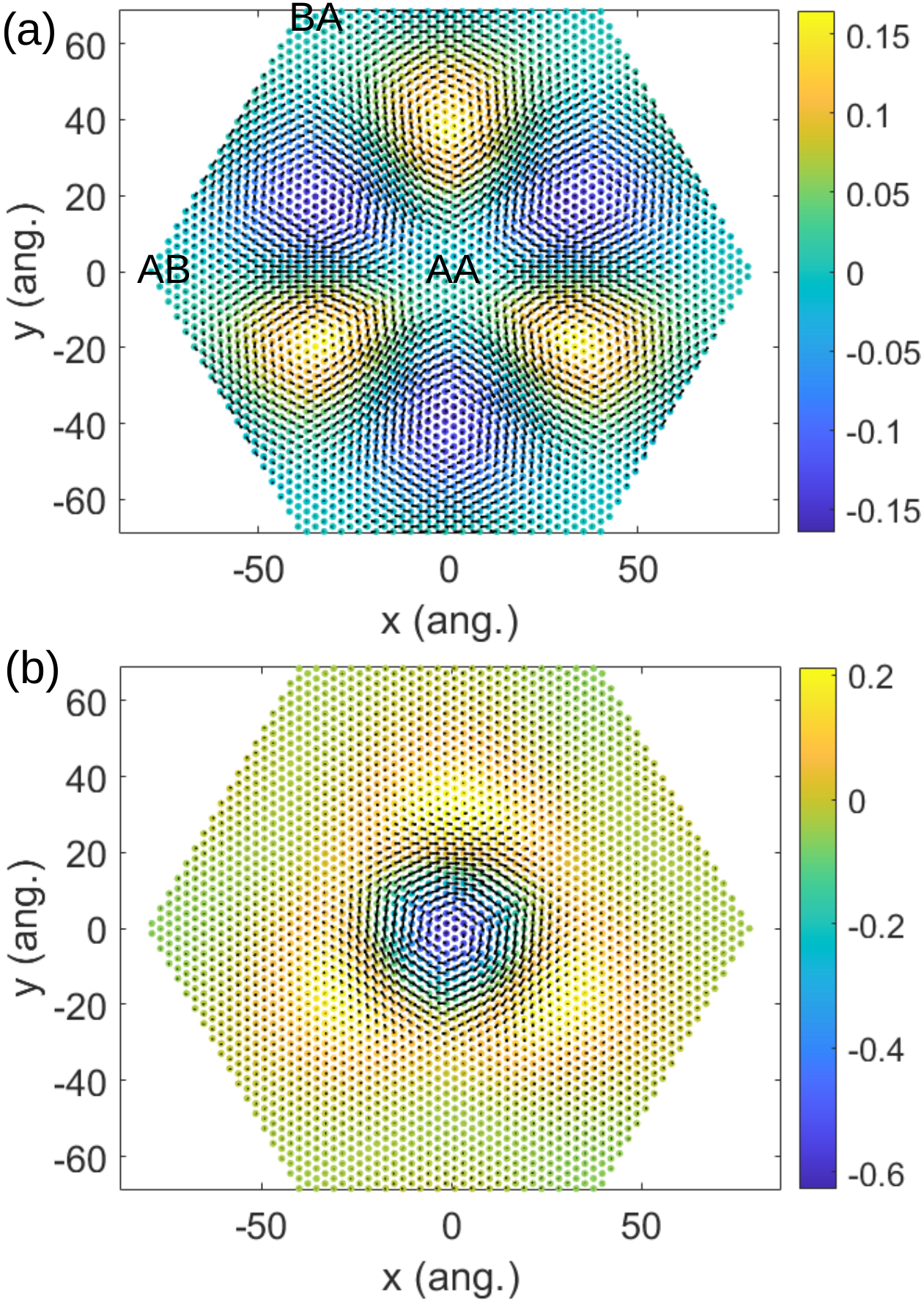}
\caption{The real-space distributions of current densities and magnetic fields for the correlated insulating states in magic-angle TBG at 1/2 filling, with $\epsilon\!=\!3.5$, $\kappa\!=\!0.005\,\angstrom^{-1}$: (a) in the $\beta$ phase (see Fig.~\ref{fig:phasediagram}(a)) with the staggered sublattice potential $\Delta\!=\!0$, and (b) in the $\alpha_M$ phase (see Fig.~\ref{fig:ahc-local}(c))  with staggered sublattice potential (induced by the aligned hBN substrate) $\Delta\!=\!15\,$meV exerted on the bottom graphene layer. The color coding indicates the strength of magnetic field in units of Gauss. The directions and amplitudes of the black arrows represent the directions and amplitudes of the current densities.} 
\label{fig:current}
\end{figure}

The phase diagrams of the magic-angle TBG at 1/2 and 0 fillings with local-exchange approximations are shown in Fig.~\ref{fig:phasediagram}(a)-(b), where the color coding indicate the indirect gaps of the HF ground states. We first consider the situation that the TBG system is not aligned with an hBN substrate such that the non-interacting Hamiltonian preserves $C_{2z}$ symmetry (see Sec.~\ref{sec:append2}). As shown in Fig.~\ref{fig:phasediagram}(a), the phase diagram at 1/2 filling consists of five different phases marked by $\alpha$, $\beta$, $\gamma$, $\delta$, and $\epsilon$ and their combinations.  These phases break different symmetries as listed in Table.~\ref{table:order-parameter}, and the corresponding symmetry-allowed order parameters in the valley-spin-sublattice space are given in Table~\ref{table:op-all}.  
There are  a few regions in the $(\kappa, \epsilon)$ parameter space in which the ground states are insulating at 1/2 filling. The insulating states can be the $\beta$, $\delta$,  $\gamma$, and $\varepsilon$ phases. To be specific, the insulating states in the $\beta$ phase take the largest area in the phase diagram. Such states are valley-polarized states with dominant order parameters $\tau_z s_{0,z}\sigma_0$, $\tau_z s_{0,z}\sigma_x$ ,which break $C_{2z}$, $\mathcal{T}$, and $C_{3z}$ symmetries, but preserves the combined symmetry operation $C_{2z}\mathcal{T}$, thus prohibits anomalous Hall effect.   By virtue of the valley polarization, the insulating state in the $\beta$ phase exhibit a remarkable real-space current pattern as shown by the black arrows in Fig.~\ref{fig:current}(a). If there are only $\tau_z s_{0,z}\sigma_0$ order parameters in the $\beta$ phase, the system would preserve both $C_{2z}\mathcal{T}$ and $C_{3z}$ symmetries on the moir\'e length scale, which give rise to six circulating  current loops in the moir\'e supercell surrounding the $AA$ region: three of them are clockwise and the other three are counter-clockwise. These compensating current loops   generate staggered magnetic fluxes and orbital magnetizations on an emergent honeycomb lattice in the moir\'e supercell.  The $\tau_z s_{0,z}\sigma_x$ order in the $\beta$ phase further breaks $C_{3z}$ symmetry,  as a result, the magnitudes of current densities flowing along the $x$ direction are slightly different from those along the $120^{\circ}$ direction (with $5\textrm{-}10\%$ difference). However, we would like to emphasize that the breaking of $C_{3z}$ symmetry is not a  necessary condition to open a gap at 1/2 filling, and does not change the nature of the correlated insulator state.
%Since such a moir\'e orbital antiferromagnetic state manifests  a threefold  current-loop pattern,  it carries nonzero toroidal octupolar moment $T_{tor}=T_{3,-3}+T_{3,3}$, where the toroidal octupolar moment in the spherical harmonic basis $T_{l,m}\sim\int_{\Omega_M} d\mathbf{r} \, r^{l}Y_{lm}^{*}(\hat{\mathbf{r}})\rho_t^{l}$. The ``toroidal charge " $\rho_t^{l}\sim -\nabla\cdot(\, 2(\mathbf{j}(\mathbf{r})\!\cdot\!\mathbf{r})\mathbf{r}-(l+3)r^2\mathbf{j}(\mathbf{r}) \,)$, with $\mathbf{j}(\mathbf{r})$ denoting the current density at position $\mathbf{r}$ within a moir\'e supercell $\Omega_M$.
Our calculations indicate that the maximal current density $\sim 1\,$nA/$\angstrom^2$, and the maximal magnetic field generated by the current loops $\sim 0.15\,$Gauss as shown by the color coding in Fig.~\ref{fig:current}(a). 

Such compensating current-loop order, or moir\'e orbital antiferromagnetic order can be interpreted as follows. Since the two flat bands for each valley of magic-angle TBG are equivalent to two degenerate zeroth pseudo Landau levels with opposite Chern numbers $\pm 1$ \cite{jpliu-prb19}, the electrons in the two pseudo LLs would perform cyclotron motions along opposite directions driven by the opposite pseudo magnetic fields, forming current loops in the moir\'e supercell flowing towards opposite directions. As discussed above, if there is only $\tau_z s_{0,z}\sigma_0$ order, the current-density distribution for each valley has to obey $C_{2z}\mathcal{T}$ and $C_{3z}$ symmetries, which gives rise to the threefold compensating current-loop pattern. The $\tau_z s_{0,z}\sigma_x$ order further breaks $C_{3z}$ symmetry which slightly distorts the threefold current pattern.
To the best of our knowledge, such moir\'e orbital antiferromagnetic  states with compensating real-space current-loop pattern  have never been experimentally observed nor theoretically discussed in any other systems. 

Since the moir\'e orbital antiferromagnetic states break $C_{2z}$ symmetry, they may exhibit giant nonlinear optical responses induced by the moir\'e antiferromagnetic order or compensating current-loop order. Generally speaking, the nonlinear photocurrent $j^{c}(\omega_c)$ is related to the time-dependent electric fields of  light via the second-order  photoconductivity: $j^{c}(\omega_c)=\sum_{ab}\sigma^{c}_{ab}(\omega_c)\,E_a(\omega_a)\,E_b(\omega_b)$, where $a, b$ denotes the spatial directions in  Cartesian coordinates, and $\omega_a$ and $\omega_b$ are the two incident light frequencies with $\omega_a+\omega_b=\omega_c$ \cite{sipe-prb00}. 
For monochromatic light, the frequency is fixed as $\pm\omega$, thus there could be two distinct second-order optical processes with $\omega_c=0$ or $\omega_c=2\omega$, corresponding to the shift-current and the second harmonic generation processes respectively \cite{sipe-prb00}. In a valley degenerate state, the nonlinear optical response is forbidden due to the presence of $C_{2z}$ symmetry. In a valley polarized state, however, the moir\'e orbital antiferromagnetic order (denoted by $\mathcal{N}_z$) can induce non-vanishing nonlinear optical responses. Therefore, the photoconductivity tensor in a valley-polarized state can be written as $\sigma^{c}_{ab}=\sigma^{c}_{ab,z}\mathcal{N}_z$. The $C_{3z}$ and $C_{2x}$ symmetries of each valley further enforce that the only non-vanishing component of the photoconductivity tensor is $\sigma^{x}_{xx}$ \cite{jpliu-tbg-optics}. In the nematic phase, however, $C_{3z}$ and $C_{2x}$ symmetries are also broken, then other components would be allowed.

\begin{figure}
\includegraphics[width=3.5in]{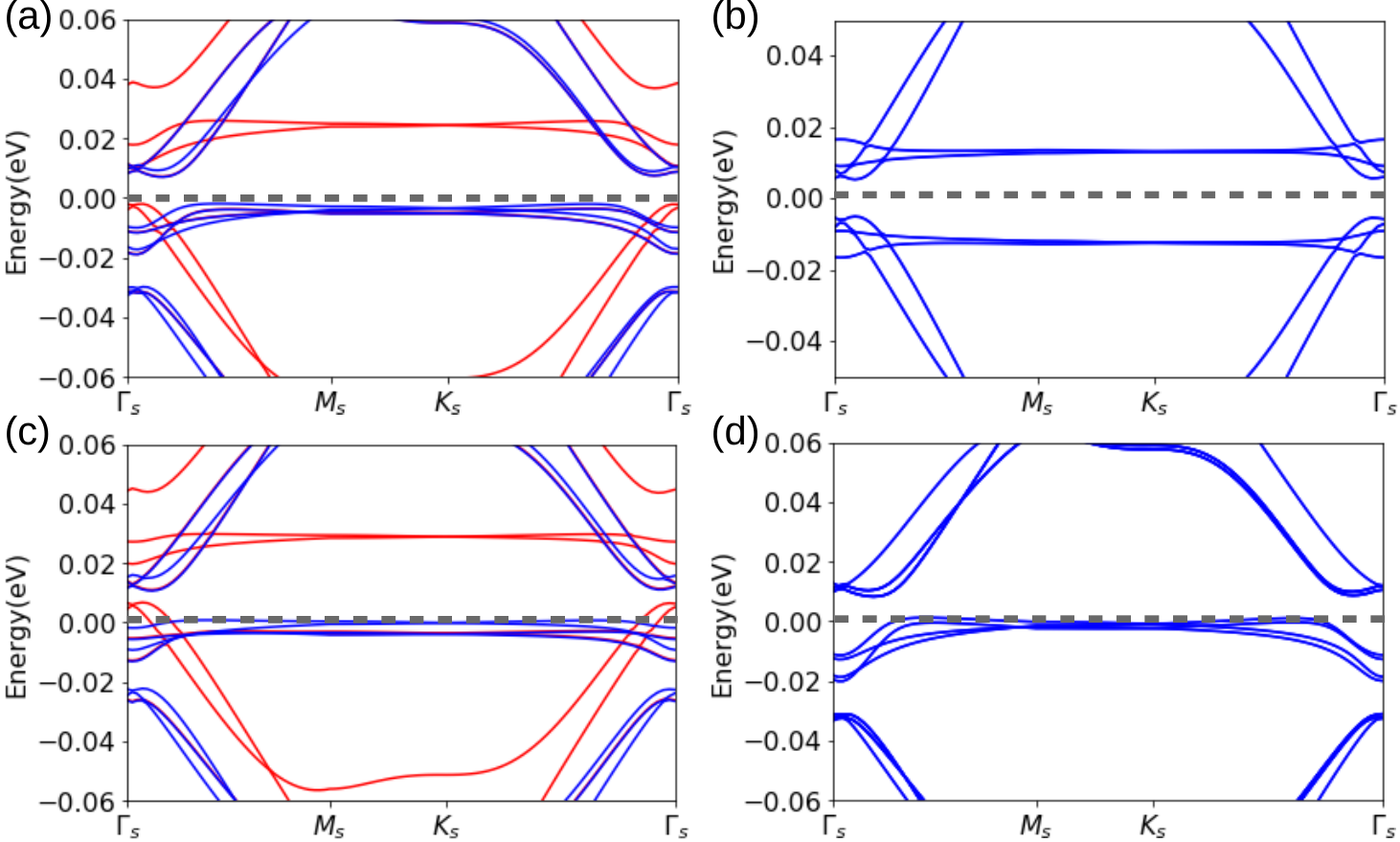}
\caption{The bandstructures of the Hartree-Fock ground states of magic-angle TBG at different fillings: (a) 1/2 filling, $\epsilon\!=\!3.5$, $\kappa\!=\!0.005\,\angstrom^{-1}$, (b) 0 filling, $\epsilon\!=\!5$, $\kappa\!=\!0.01\,\angstrom^{-1}$, (c) 1/4 filling, $\epsilon\!=\!3.5$, $\kappa\!=\!0.005\,\angstrom^{-1}$, (d) 3/4 filling, $\epsilon\!=\!5$, $\kappa\!=\!0.01\,\angstrom^{-1}$. The blue and red lines in (a) and (c) represent the majority and minority spins. The gray dashed lines mark the chemical potentials.} 
\label{fig:bands}
\end{figure}

Moreover, although the spin rotational symmetry is also broken in the $\beta$ phase,  the spin character of the valence-band maximum (conduction-band minimum) of the resulted  Hartree-Fock bandstructures  is  opposite to the majority (minority) spin of the system (see Fig.~\ref{fig:bands}(a)). As a result, despite the non-vanishing spin splittings, the energy gaps in the $\beta$ phase should still decrease  in response to external magnetic field along any direction due to the spin Zeeman effect. Such ``spin-paramagnetic-like" behavior is consistent with experimental observations \cite{cao-nature18-mott}. %The dominant order parameters in the $\beta$ insulating phase at 1/2 filling are given in Table~\ref{table:op-half}.

On the other hand, the insulating states in the $\delta +\alpha$ phase in the lower left corner of Fig.~\ref{fig:phasediagram}(a) include both the IVC order ($\delta$) and VP order ($\alpha$). In particular, the $\delta$ phase breaks the $U_v(1)$ and $SU_v(2)$ symmetry, while the $\alpha$ phase is a valley polarized phase which breaks $C_{2z}$ and $\mathcal{T}$ symmetries. 
In the  $\gamma +\alpha$ phase, the system further  breaks global spin rotational symmetry based on the $\delta+\alpha$ phase.
The insulating states in the $\varepsilon$ phase are spin antiferromagnetic states with opposite spin magnetizations on the $A$ and $B$ sublattices. All the symmetry-allowed order parameters in the $\beta$, $\gamma$,  $\delta$, and $\varepsilon$ phases are listed in Table~\ref{table:op-all}. 
When both $\epsilon$ and $\kappa$ are large, the Coulomb interaction is not strong enough to open a global gap, and the system stays in a valley polarized metallic state labeled by $\alpha$ phase in Fig.~\ref{fig:phasediagram}(a), which breaks both $C_{2z}$ and $\mathcal{T}$, but preserves the combined symmetry $C_{2z}\mathcal{T}$. %As discussed above, 
Given that the insulating states in the $\beta$ phase takes the largest area of the phase diagram and that the properties of such states are consistent with experimental observations
, we propose that the experimentally observed correlated insulating states at $\pm 1/2$ fillings are the valley-polarized, spin-splitted states exhibiting such moir\'e orbital antiferromagnetic order (compensating current-loop order). Such a valley-polarized correlated insulator state in the $\beta$ phase becomes even more robust after including the effects of nonlocal exchange interactions as will be discussed in detail in Sec.~\ref{sec:nonlocal-exchange}.

At the charge neutrality point (0 filling), with local-exchange approximation the system stays in a valley-polarized and spin paramagnetic insulating state for most of the interaction parameters ($\epsilon$, $\kappa$) ($\alpha$ phase in Fig.~\ref{fig:phasediagram}(b)), in which the gap  varies from $1\,$meV to $13\,$meV for different ($\epsilon$, $\kappa$) parameters. In particular, for physically reasonable dielectric constant $\epsilon\sim 3\textrm{-}5$, and $\kappa\sim 0.005\,\angstrom^{-1}$, the calculated gap in the $\alpha$ phase $\sim 1\textrm{-}5\,$meV, which is in qualitative agreement with the measured transport gap $\sim 0.86\,$meV \cite{efetov-nature19}.
As shown in Table~\ref{table:order-parameter}, the $\alpha$ phase breaks both $C_{2z}$ and $\mathcal{T}$ symmetries, but preserve the combined $C_{2z}\mathcal{T}$ symmetry, which prohibits anomalous Hall effect, but may exhibit nonlinear optical responses as discussed above. The insulating states in the $\alpha$ phase are also orbital antiferromagnetic states with opposite circulating currents and staggered magnetic fluxes on the moir\'e length scale. The current-density pattern and the distributions of the magnetic fields are similar to that at 1/2 filling  as shown in Fig.~\ref{fig:current}(a).
The dominant order parameters for the correlated insulator in the $\alpha$ phase at the charge neutrality point are $\tau_z s_0\sigma_0$ and $\tau_z s_0\sigma_x$. %as shown in Table~\ref{table:op-half}. 
At 1/4 filling, the system mostly stays in a metallic $\beta$ phase as shown in Fig.~\ref{fig:phasediagram}(c), which breaks global spin $SU(2)$, and preserves $C_{2z}\mathcal{T}$ symmetry.  At 3/4 filling,  without the alignment of the hBN substrate, the system remains as a valley polarized metal ($\alpha$ phase) preserving $C_{2z}\mathcal{T}$ symmetry  for most of the interaction parameters as shown in Fig.~\ref{fig:phasediagram}(d).

The bandstructures of the HF ground states with local-exchange approximations at different integer fillings are shown in Fig.~\ref{fig:bands}. In Fig.~\ref{fig:bands}(a), as discussed above, we show the HF bandstructure at 1/2 filling with $\epsilon\!=\!3.5$, $\kappa\!=\!0.005\,\angstrom^{-1}$, where the blue and red lines denote the majority and minority spin spices. As discussed above, although there are non-vanishing spin splittings, the spin character of the valence-band maximum (conduction-band minimum) of the   Hartree-Fock bandstructure  is  opposite to the majority (minority) spin of the system . As a result, the gap still decreases in response to external magnetic field due to spin Zeeman effect as shown in the inset of Fig.~\ref{fig:phasediagram}(a). The HF bandstructure  at 0 filling with $\epsilon\!=\!5$, $\kappa\!=\!0.01\,\angstrom^{-1}$ is shown in Fig.~\ref{fig:bands}(b). The valley-polarized order parameters $\tau_z s_0\sigma_0$ and $\tau_z s_0\sigma_x$ open a  gap at the charge neutrality point, leading to an insulating state $\sim 10\,$meV, which breaks both $C_{2z}$ and $\mathcal{T}$ symmetries, but preserve the $C_{2z}\mathcal{T}$ symmetry. In Fig.~\ref{fig:bands}(c) we show the spin-resolved HF bandstructure at 1/4 filling with $\epsilon\!=\!3.5$, $\kappa\!=\!0.005\,\angstrom^{-1}$. The system is in a valley and spin-split metallic state.  In Fig.~\ref{fig:bands}(d) we show the HF bandstructure at 3/4 filling with $\epsilon\!=\!5$, $\kappa\!=\!0.01\,\angstrom^{-1}$, in which the bands are valley split but the spin degeneracy is preserved.

\subsection{Quantum anomalous Hall effects in hBN-aligned TBG}

\label{sec:local-exchange-b}

We continue to study the magic-angle TBG system aligned with hBN substrate at different integer fillings in the local-exchange approximation.  When hBN is aligned with TBG, the boron (nitrogen) atom is below the $A (B)$ sublattice of the bottom-layer graphene, which imposes a staggered sublattice potential $\Delta\!\approx\!15\,$meV on the bottom layer graphene. 
As a consequence, each valley of TBG no longer preserves the $C_{2z}\mathcal{T}$ symmetry, and the anomalous Hall effect and  orbital ferromagnetism are expected to show up once the system becomes valley polarized. Once such a staggered sublattice potential (also known as the ``Dirac mass term") is included (see Appendix ~\ref{sec:append2}), our calculations indicate that the valley-polarized states are energetically favored over the IVC states for most of the $(\kappa, \epsilon)$ parameters at all integer fillings.

\begin{figure}
\includegraphics[width=3.5in]{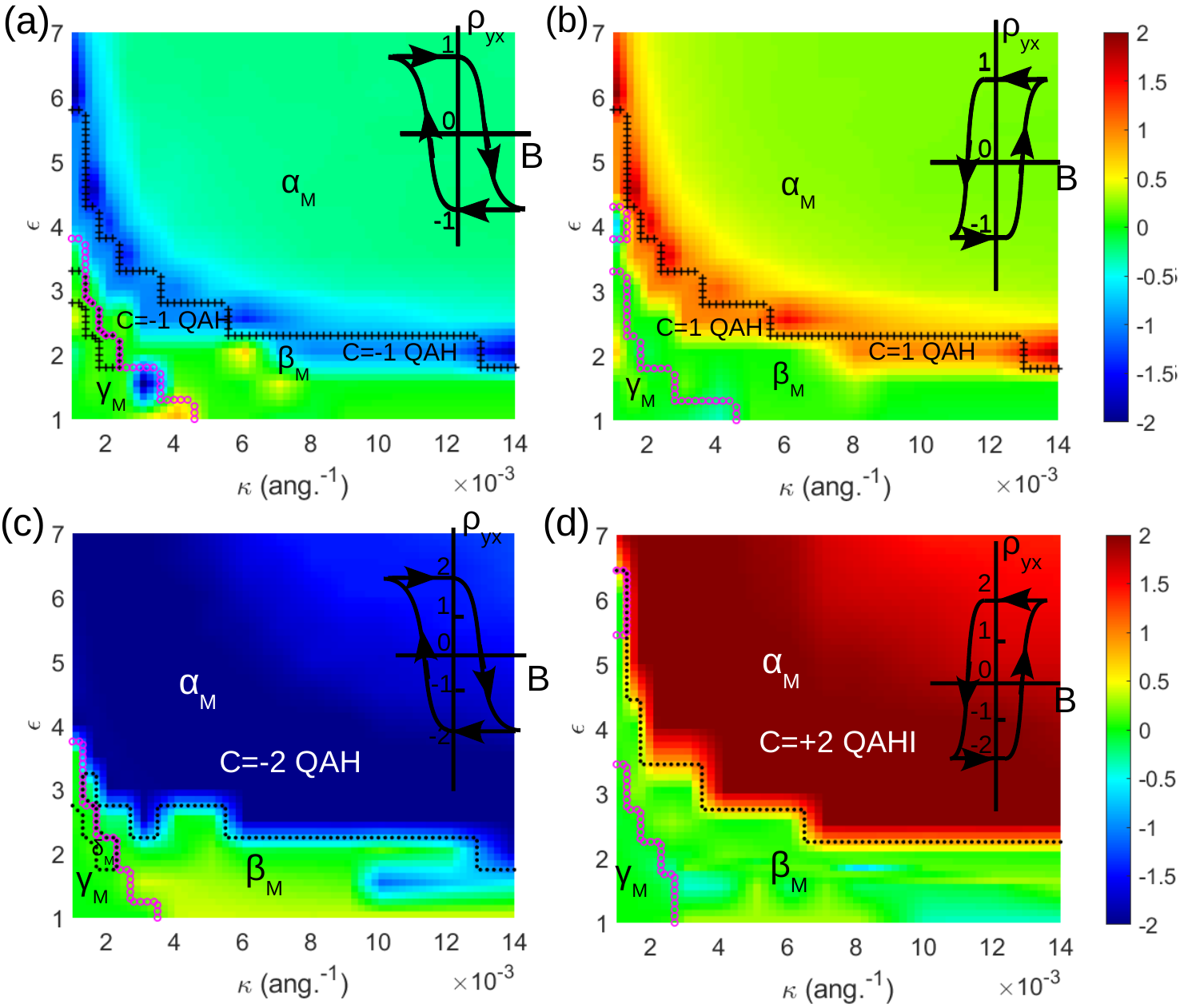}
\caption{The calculated anomalous Hall conductivities (in units of $e^2/h$) of the Hartree-Fock ground states for TBG aligned with hBN substrate with local-exchange approximation: (a) at $3/4$ filling, (b) at $-3/4$ filling, (c) at 1/2 filling, and (d) at -1/2 filling. The insets schematically show the hysteresis loops of the QAH effects at different fillings.}
\label{fig:ahc-local}
\end{figure}

Quantum anomalous Hall effects with the Chern number $\pm 1$ has been experimentally observed at $3/4$ filling of the flat bands in hBN-aligned TBG around the magic angle \cite{young-tbg-science19}. 
In order to understand such intriguing QAH effect at $3/4$ filling and the crucial role of the alignment of the hBN substrate \cite{sharpe-science-19, young-tbg-science19}, we have calculated the HF ground states at $\pm 3/4$ filling of the flat bands in magic-angle TBG with local-exchange approximation. When the staggered sublattice potential  $\Delta\!=\!0$, the HF ground states at $3/4$ filling are mostly the valley-polarized and spin degenerate metallic states in the $\alpha$ phase (see Fig.~\ref{fig:phasediagram}(d)) with vanishing AHC, which is prohibited by $C_{2z}\mathcal{T}$ symmetry. The situations are drastically different when an hBN substrate is aligned with the TBG system. When the staggered sublattice potential on the bottom layer graphene $\Delta\!=\!15\,$meV, the calculated AHC of the HF ground states (with local-exchange approximation)  at $3/4$ and $-3/4$ fillings are presented in Fig.~\ref{fig:ahc-local}(a) and (b) respectively. In addition to the valley-split metallic phase ($\alpha_M$ phase), for some intermediate interaction parameters ($\kappa$, $\epsilon$) the ground state can be a  QAH insulator with the Chern number $\mp1$ as marked by ``$C=-1$ QAH" and ``$C=1$ QAH" in Fig.~\ref{fig:ahc-local}(a) and (b). Such QAH states emerge in the $\beta_M$ phase  when both $\mathcal{T}$ and global spin $SU(2)$ symmetry are spontaneously broken, i.e., when the system exhibits both orbital and spin ferromagnetic ordering. 
It is  interesting to note that given the same valley polarization, the orbital magnetizations in the $C\!=\!-1$ and $C\!=\!1$ QAH insulating states at $3/4$ and $-3/4$ fillings have the \textit{same sign} ( both $\sim 5\hbox{-}10\,\mu_{\textrm{B}}$  per moir\'e supercell). It follows that if the orbital  magnetizations at $\pm 3/4$ fillings are both polarized along $+z$ direction by external magnetic fields, the AHCs at the opposite fillings would be opposite in sign, thus the system would exhibit  hysteresis loops with exactly \textit{opposite chiralities}  at $-3/4$ and $3/4$ fillings as schematically illustrated in the insets of Fig.~\ref{fig:ahc-local}(a) and (b). The opposite hysteresis loops at electron and hole fillings would be an important experimental signature for the QAH phenomena in TBG. 
Including the nonlocalness of the exchange interactions would significantly enhance the spin polarization, which greatly enlarges the area of the $C=\mp 1$ QAH phase in the ($\kappa$, $\epsilon$) parameter space, which will be discussed in detail in the following section.

\begin{figure*}
\includegraphics[width=4.8in]{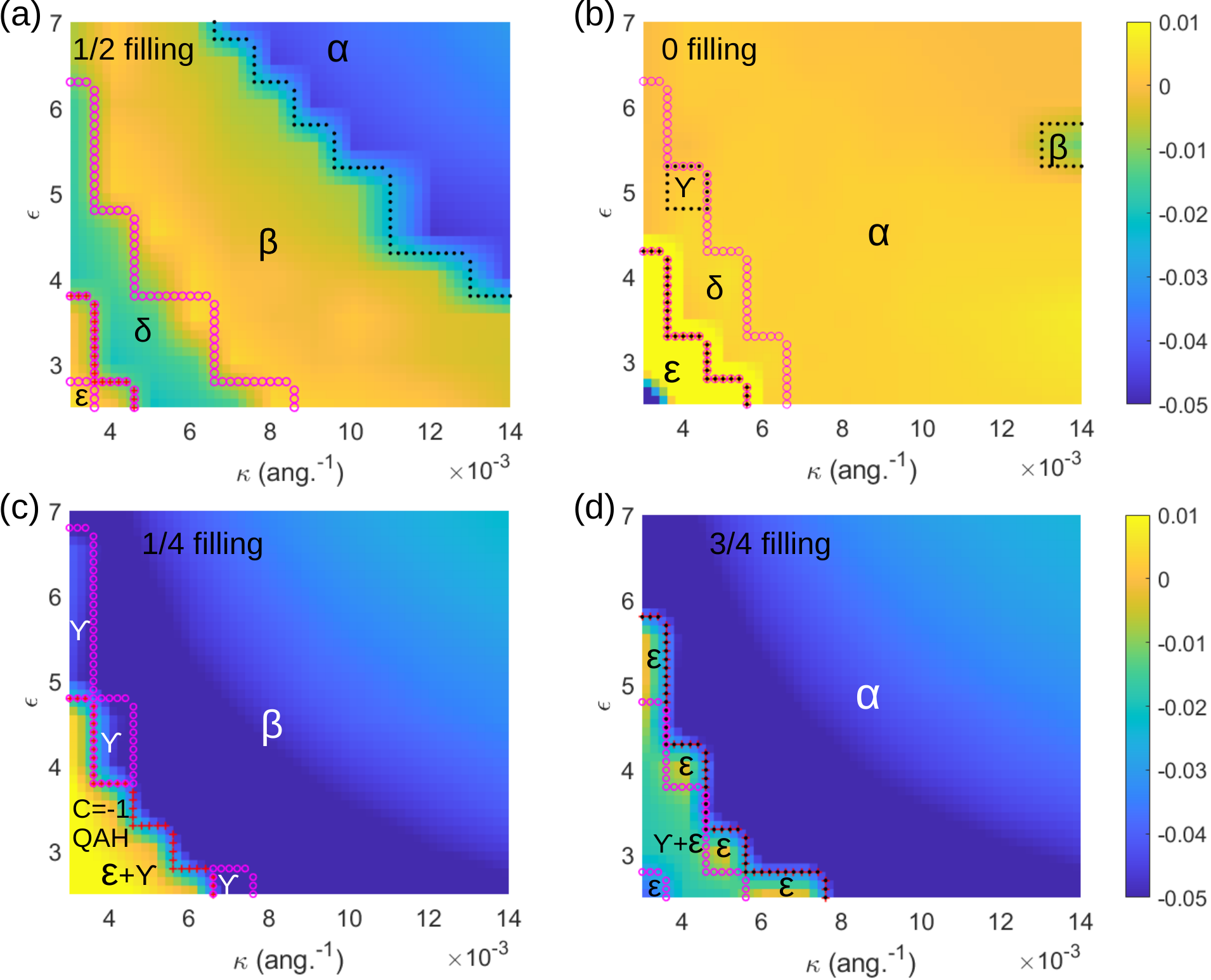}
\caption{The indirect gaps (in units of eV) of the Hartree-Fock ground states with nonlocal exchange interactions for magic-angle TBG (without hBN alignment): (a) at 1/2 filling,  (b) at 0 filling, (c) at 1/4 filling, (d) at 3/4 filling.} 
\label{fig:phasediagram-nonlocal}
\end{figure*}

In Fig.~\ref{fig:ahc-local}(c) and (d) we show the  calculated AHC (in units of $e^2/h$) of the HF ground states at 1/2  and -1/2 fillings of the hBN-aligned TBG with local-exchange approximation.  The ground states  are  $C\!=\!\mp 2$ QAH insulating states at $\pm 1/2$ filling for intermediate interaction strengths.  When the interaction is weak, i.e., with large $\epsilon$ and large $\kappa$, the system becomes a valley-splitted ``Chern metal" characterized by large but non-quantized AHC, as shown in the upper-right corner of Fig.~\ref{fig:ahc-local}(b).  Both the $C=\pm 2$ QAH states and the Chern-metal states are in the $\alpha_M$ phase, which spontaneously breaks $\mathcal{T}$ symmetry. Moreover, since $C_{2z}$ is already broken in the non-interacting Hamiltonian of hBN-aligned TBG,  $C_{2z}\mathcal{T}$ symmetry is also broken, giving rise to the QAH effects in the valley polarized states.  On the other hand, the $\alpha_M$ phase preserves all the continuous symmetries including the spin $SU(2)$ symmetry (see Table~\ref{table:order-parameter}). It follows that the $C\!=\!\pm2$ QAH states in the $\alpha_M$ phase are \textit{orbital ferromagnetic} and \textit{spin paramagnetic} states. Such states exhibit large orbital magnetization $\sim 5-10\,\mu_{\textrm{B}}$ per moir\'e supercell, which can be stablized by out-of-plane magnetic fields by virtue of the orbital magnetic Zeeman effect, but could be strongly suppressed by in-plane magnetic field due to the (interaction-enhanced) spin Zeeman effect \cite{supp_info}. The  $C\!=\!\mp 2$ QAH states at $\pm 1/2$ fillings also exhibit hysteresis loops with opposite chiralities as discussed above.
The orbital ferromagnetic states  exhibit chiral current loops  circulating in the $AA$ region as shown in Fig.~\ref{fig:current}(b), which generate magnetic fields peaked in the $AA$ region, with the maximal amplitude $\sim 0.6\,$Gauss. 
It should be noted again that the results presented in Fig.~\ref{fig:ahc-local}(b) are  based on local-exchange approximation. Including the nonlocalness of the exchange interaction would significantly enhance the spin polarization of the system, such that a topologically trivial insulating state with both valley and spin polarizations would dominate over the $C=\pm 2$ QAH state at 1/2 filling.  More details will be discussed in Sec.~\ref{sec:nonlocal-exchange}.

\section{Phase diagrams with nonlocal exchange interactions}
\label{sec:nonlocal-exchange}

Having established the phase diagrams at various fillings with local-exchange approximations, we continue to study the effects of nonlocal exchange interactions.
Since the nonlocal Fock operator $O^{F}_{\gamma\gamma'}(\mathbf{r})$ correlates two electrons separated by a finite displacement vector $\mathbf{r}$, its Fourier transform $O^{F}_{\gamma\gamma'}(\mathbf{k})=\int d\mathbf{r}\,e^{i\k\cdot\mathbf{r}}\,O^{F}_{\gamma\gamma'}(\mathbf{r})$ acquires nontrivial $\k$ dependence. Therefore, in principle, the Fock order parameters at each $\k$ point has to be treated as independent variational parameters, which eventually yields more than ten thousands independent variational parameters for each calculation. This is beyond our computational capability. In order to resolve this problem, we assume the Fock order parameters $O^{F}_{\gamma\gamma'}(\mathbf{k})$ have some simple analytic dependence on $\k$. In particular, we assume $O^{F}_{\gamma\gamma'}(\mathbf{k})$ decays in a  Gaussian form with respect to $\vert\k\vert$, i.e.,
\begin{equation}
O^{F}_{\gamma\gamma'}(\mathbf{k})=O^{F,(0)}_{\gamma\gamma'}\,e^{-\kappa_F\,\vert\k\vert^2 L_s^2}
\end{equation}
where $O^{F,(0)}_{\gamma\gamma'}$ represents the $\k$ independent Fock order parameters in the valley, spin, sublattice space, $L_s$ is the moir\'e lattice constant, and $\kappa_F$ is a dimensionless parameter characterizing the spread of the Fock order parameters in $\k$ space. $\kappa_F$ will be treated as an extra free parameter in the nonlocal-exchange Hartree-Fock calculations to be determined variationally.

\subsection{Correlated insulators at zero and half fillings in TBG without hBN alignment}

We first consider the correlated insulator phases in TBG without hBN alignment. In Fig.~\ref{fig:phasediagram-nonlocal} we show the indirect gaps of the HF ground states including nonlocal exchange interactions at (a) 1/2, (b) 0, (c) 1/4, and (d) 3/4 fillings respectively. Comparing Fig.~\ref{fig:phasediagram-nonlocal} with Fig.~\ref{fig:phasediagram}, we see that  the phase diagrams with nonlocal and local exchange interactions are qualitatively consistent with each other. A minor difference is that the valley polarized correlated insulators in the $\beta$ phase at 1/2 filling  and in the $\alpha$ phase at 0 filling occupy larger areas in the ($\kappa$, $\epsilon$) parameter space in the nonlocal HF calculations. Therefore, the conclusions drawn in Sec.~\ref{sec:local-exchange} that the nature of the correlated insulators at 0 and 1/2 fillings are valley polarized states with compensating moir\'e current-loop order are reinforced by the HF calculations including nonlocal exchange interactions. The other minor difference between the local and nonlocal exchange calculations is that the nonlocal exchange tends to enhance the bandwidth of the low-energy bands due to the exchange-hole effect: two electrons with the same flavor degrees of freedom have to avoid meeting with each other in real space due to Pauli exclusion principle, which effectively amplify the kinetic energy. As a result, the calculated indirect gaps of the correlated insulator phases including nonlocal exchange are smaller than those obtained  from local-exchange calculations. 

The correlated insulator phase at 1/2 filling ($\beta$ phase) exhibit both valley and spin polarizations. However, by virtue of the exchange-hole effects, the low-energy flat bands become more dispersive such that the spin character of the VBM near $\Gamma_s$ are opposite to the majority spin. %This is clearly shown in Fig.~\ref{fig:bands-nonlocal}(a), in which we plot the  HF bandstructures with nonlocal exchange interaction at 1/2 filling with $\epsilon\!=\!4.5$, $\kappa\!=\!0.005\,\angstrom^{-1}$, where the blue (red) lines denote energy bands from majority (minority) spin species.
As a result, despite the nonzero spin polarization, the energy gap of the HF ground state still decreases linearly in response to external magnetic field due to spin Zeeman effect, which is consistent with experimental observations \cite{cao-nature18-mott}. 
On the other hand, the correlated insulator at the CNP in the $\alpha$ phase are valley polarized and spin degenerate states.  The correlated insulators at both 1/2 filling and the CNP exhibit the compensating current-loop pattern and staggered orbital magnetic fluxes in the moir\'e supercell as shown in Fig.~\ref{fig:current}(a), which are expected to exhibit nonlinear optical responses as discussed in Sec.~\ref{sec:local-exchange-a}. %The spin-degenerate HF bandstructures with nonlocal exchange interactions at the CNP with $\epsilon\!=\!4.5$, $\kappa\!=\!0.005\,\angstrom^{-1}$ are shown in Fig.~\ref{fig:bands-nonlocal}(b).

\begin{figure*}
\includegraphics[width=5.5in]{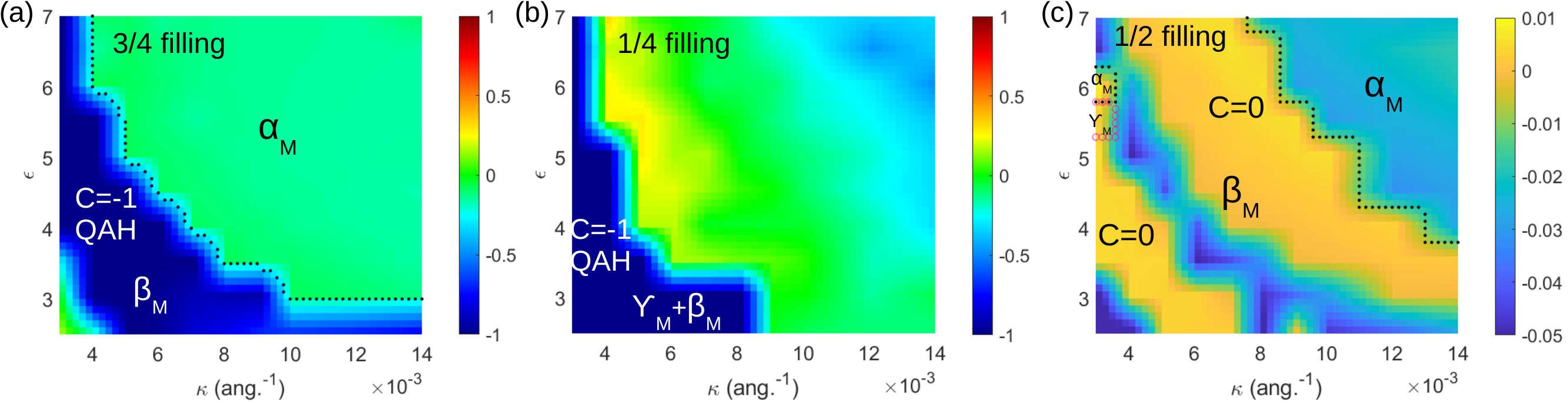}
\caption{The anomalous Hall conductivities (in units of $e^2/h$) and indirect gap (in units of eV) of the Hartree-Fock ground states with nonlocal exchange interactions for hBN-aligned TBG at the magic-angle: (a) AHC at 3/4 filling,  (b) AHC at 1/4 filling, and (c) indirect gap  at 1/2 filling.} 
\label{fig:ahcgap-nonlocal}
\end{figure*}

At 1/4 and 3/4 fillings, the HF ground states with nonlocal exchange interactions shown in Fig.~\ref{fig:phasediagram-nonlocal}(c)-(d) are also metallic $\beta$ and $\alpha$ phases for most of the interaction parameters ($\kappa$, $\epsilon$),  which are consistent with those calculated by local-exchange approximation as shown in Fig.~\ref{fig:phasediagram}(c)-(d).  For strong Coulomb interactions, at 1/4 filling the $C_{2z}\mathcal{T}$ symmetry can be broken spontaneously, leading to a $C=\pm 1$ QAH insulating state with co-existing VP-type and IVC-type order parameters, with the VP-type order being dominant.  This is marked as  in the $\varepsilon+\gamma$ phase as shown in the lower-left corner of Fig.~\ref{fig:phasediagram-nonlocal}(c) %
\footnote{We note that a $C\!=\!\pm 1$ QAH state at 1/4 filling without hBN alignment has been experimentally observed  recently \cite{efetov-tbg-supercond-chern-arxiv20} after the submission of our manuscript. Our results at 1/4 filling for strong Coulomb interactions are also consistent with those obtained by Lian \textit{et al.} based on perturbation theories in the strong-coupling limit \cite{lian-tbg-iv-arxiv20}.  }. 
At 3/4 filling, without the staggered sublattice potential, the $C_{2z}\mathcal{T}$-broken $C=\pm 1$ QAH phase is  a metastable phase for strong Coulomb interactions, but  its energy is typically higher than that of the other phase with coexisting IVC-type and VP-type order parameters dubbed as ``$\gamma+\varepsilon$“ phase in the  lower-left corner of Fig.~\ref{fig:ahcgap-nonlocal}(d). The co-existing $\gamma+\varepsilon$ phase turns out to be either metallic or insulating with zero Chern number, thus the $C=\pm 1$ QAH state is unstable for TBG at 3/4 filling without hBN alignment. This is consistent with the fact that the $C=\pm1$ QAH effect can be observed at 3/4 filling only when the hBN substrate becomes aligned with the TBG sample \cite{sharpe-science-19, young-tbg-science19}.
It is worthwhile to note that although  resistivity peak (which decreases with temperature) has been observed at  3/4 fillings in transport measurements \cite{efetov-nature19}, it is not necessarily associated with a correlated insulator. Instead, a  resistivity peak can also be contributed by van Hove singularity due to the large density of states and reduced Fermi velocity.  Detailed Hall density analysis reveals that the resistivity peak at 3/4 filling is actually contributed by van Hove singularity without gap opening \cite{eva-tbg-chern-arxiv20}.

\subsection{Quantum anomalous Hall states in hBN-aligned TBG}

We continue to study the correlated insulator phases and quantum anomalous Hall effects in hBN-aligned TBG. In Fig.~\ref{fig:ahcgap-nonlocal}(a) and (b) we show the AHC of the Hartree-Fock ground states including nonlocal exchange effects at 3/4 and 1/4 fillings respectively. At 3/4 filling, the area of the QAH phase with Chern number -1 is significantly extended compared with that from local-exchange calculations (compare Fig.~\ref{fig:ahc-local}(a) with Fig.~\ref{fig:ahc-local}(a)). It turns out that  the spin polarization has been significantly enhanced by the nonlocal exchange interactions, which is essential in opening a topologically nontrivial gap at 3/4 filling in hBN-aligned TBG.   The QAH state at 3/4 filling is a orbital and spin ferromagnetic state with dominant order parameters $\tau_0 s_0\sigma_j$, $\tau_0 s_z\sigma_j$, $\tau_z s_0\sigma_j$, and $\tau_z s_z\sigma_j$ ($j=0, x, z$).  
At 1/4 filling, surprisingly, we find that for relatively strong Coulomb interactions (lower left corner of Fig.~\ref{fig:ahcgap-nonlocal}(b)), the ground state is also a $C=-1$ QAH state with co-existing $\gamma_{M}$ and $\beta_M$ phases.  Such a state is different from that at 3/4 filling in the sense that it is a state with co-existing  VP-type and IVC-type order parameters. Both types of order parameters are crucial in opening a topologically nontrivial gap.
This implies that QAH effect with Chern number $\pm 1$ could be observed at 1/4 filling of hBN-aligned as well. %The  HF bandstructures at 1/4 and 3/4 fillings for hBN-aligned TBG with $\epsilon\!=\!4.5$ and $\kappa\!=\!0.004\,\angstrom^{-1}$ are shown in Fig.~\ref{fig:bands-nonlocal}(c) and (d) respectively, where the blue (red) lines represent the majority (minority) spin species.

At 1/2 filling, the HF ground states with nonlocal exchange interactions are drastically different from those obtained with local-exchange approximation. In Fig.~\ref{fig:ahcgap-nonlocal}(c) we show the indirect gaps of the HF ground states with nonlocal exchange interactions at 1/2 filling of hBN-aligned TBG. For intermediate and/or strong Coulomb interactions, the system stays in correlated insulating states in the $\beta_M$ phase with zero Chern number. The $C=-2$ QAH states obtained from local-exchange approximation (Fig.~\ref{fig:ahc-local}(b)) are suppressed by the large spin polarizations induced by the nonlocal exchange interactions. In other words, at 1/2 filling the system undergoes a transition from a valley polarized and spin degenerate QAH insulator to a valley and spin polarized trivial insulator due to the inclusion of nonlocal exchange effects. Experimentally QAH effect with Chern number -1 has been observed at 3/4 filling in hBN-aligned TBG around the magic angle, but no anomalous Hall signal has been observed at 1/2 filling \cite{young-tbg-science19, sharpe-science-19}. This is fully consistent with our Hartree-Fock calculations including nonlocal exchange interactions. By virtue of an approximate particle-hole symmetry for the low-energy bands of TBG, we have only calculated the phase diagrams including nonlocal exchange interactions at electron fillings. The phase diagrams at the hole fillings are similar to those at the electron fillings with opposite anomalous Hall conductivities and Chern numbers (if any).

\section{Discussions}

\begin{table}[bth]
\caption{Correlated  phases for TBG without hBN alignment with $\epsilon\!=\!4.5$ and $\kappa\!=\!0.005\,\angstrom^{-1}$}
\begin{ruledtabular}
\begin{tabular}{lclclclc}
filling & -3/4 &-1/2 & -1/4 & 0  & 1/4 & 1/2 & 3/4 \;\\
%$\epsilon$ & 4.5 & 4.5 & 4.5 & 4.5 & 4.5 & 4.5 & 4.5 \;\\
%$\kappa$ ($\angstrom^{-1}$) & 0.005 & 0.005 & 0.005 & 0.005 & 0.005 & 0.005 & 0.005\;\\
phase & $\alpha$ & $\beta$  &  $\alpha$ & $\alpha$  &  $\alpha$ &  $\beta$ & $\alpha$\;\\
I/M & M & I & M & I & M & I & M  \footnote{``I" for insulator, ``M" for metal}\\
%AHC ($e^2/h$) &  0  & 0 & 0 & 0 & 0 & 0 & 0 
\end{tabular}
\end{ruledtabular}
\label{table:phase1}
\end{table}

In the actual TBG system encapsulated by hBN, the effective dielectric constant is  induced by the surrounding hBN with $\epsilon\sim 4\textrm{-}5$ .  It is difficult to determine the actual screening length,  but according to our nonlocal-exchange HF calculations it seems that a reasonable value which successfully reproduces the experimental phenomena at most of the integer fillings  is $\kappa\sim 0.004\textrm{-}0.005\,\angstrom^{-1}$. In Table~\ref{table:phase1}  we identify the correlated phases  at each  integer filling for TBG without hBN alignment. We find that with $\epsilon\!=\!4.5$ and $\kappa\!=\!0.005\,\angstrom^{-1}$,  the correlated insulator states at most fillings can be  explained very well.  Recently, a cascade of phase transitions have been observed at all integer fillings of TBG at the magic angle \cite{yazdani-cascade-tbg,ilani-cascade-tbg}. STM and compressibility measurements show that  at integer fillings of the flat bands, the electrons tend to fully occupy one of the fourfold degenerate flat bands, lifting the flavor (valley and spin) degeneracy. These observations are fully consistent with our calculations presented above.

The correlating insulating states and the quantum anomalous Hall phenomena  presented in this paper are all at the magic angle $\theta\!=\!1.05^{\circ}$. Indeed these phenomena are very sensitive to the twist angle. Our calculations indicate that both the correlated insulating states and the QAH effects discussed above appear \textit{only at the magic angle}. If the twist angle deviates from the magic angle, the  bandwidth would be significantly enhanced. As a result, for the same interaction parameters $\epsilon$ and $\kappa$, the Coulomb interactions are no longer strong enough to open a global gap, and the system becomes metallic. We refer the readers to Supplemental Material for more details about the twist-angle dependence of the correlated insulating states and the QAH effects.

To conclude, using a generic unrestricted moir\'e Hartree-Fock variational method, we have successfully explained  the correlated insulating states and the QAH phenomena at most integer fillings in the TBG system. We propose that the correlated insulating states observed at $\pm 1/2$ fillings and at the charge neutrality point are  the valley polarized insulating states which break both $C_{2z}$ and $\mathcal{T}$ symmetries, but preserve $C_{2z}\mathcal{T}$ symmetry. Such correlated insulating states are moir\'e orbital antiferromagnetic states exhibiting opposite circulating current loops on an emergent  honeycomb lattice in the moir\'e supercell. Within the same theoretical framework, we have also successfully explained  the $C\!=\!\pm1$ QAH effects at $3/4$ fillings.  The QAH states at electron and hole fillings exhibit hysteresis loops with opposite chiralities, which is a unique feature for the QAH effects in TBG.
We also predict the possible $C=\mp 1$ QAH state at $\pm 1/4$ fillings and $C=\mp 2$ QAH states at $\pm 1/2$ fillings in hBN-aligned TBG. Our work is a significant step forward in understanding the correlation effects and band topology in twisted bilayer graphene, and may provide useful guidelines for future experimental and theoretical studies.

\acknowledgements
J.L. and  X.D. acknowledge  financial support from the Hong Kong Research Grants Council (Project No. GRF16300918) and the start-up grant of ShanghaiTech University. We thank Lei Wang, Hongming Weng, Andrei Bernevig,  Gang Chen, and Yang Zhang for helpful discussions.

\appendix

\section{The moir\'e superlattice structure}
\label{sec:append1}
As discussed in the main text, the commensurate moir\'e pattern is formed when the top-layer graphene is rotated with respect to the bottom layer by  the  angle $\theta(m)$, where $m$ is an integer obeying the condition $\cos{\theta(m)}=(3m^2+3m+1/2)/(3m^2+3m+1)$ \cite{castro-neto-prb12}. There are periodic modulations of the $AA$, $AB$ and $BA$ regions in the moir\'e pattern. In the $AA$ region, the $A$ ($B$) sublattice of the top layer is roughly on top of the same sublattice of the bottom layer; while in the $AB$ ($BA$) regions, the $B (A)$ sublattice of the top layer is roughly on top of the $A (B)$ sublattice of the bottom layer. The  lattice vectors of the moire superlattice can be chosen as $\mathbf{t}_1=[\sqrt{3}L_s/2,L_s/2]$, and $\mathbf{t}_2=[\sqrt{3}L_s/2,-L_s/2]$, where $L_s=\vert\mathbf{t}_1\vert=a/(2\sin{\theta/2})$ is  the superlattice constant of the moire supercell, and $a=2.46$\,\angstrom is the lattice constant of monolayer graphene.  The corresponding moir\'e reciprocal lattice vectors $\mathbf{g}_1\!=\![2\pi/(\sqrt{3}L_s), 2\pi/L_s]$, and $\mathbf{g}_1\!=\![2\pi/(\sqrt{3}L_s), -2\pi/L_s]$.
After the twist, the $K$ ($K'$) points of the two monolayers $K_1$ ($K_2'$) and $K_2$ ($K_1'$) are mapped to the $K_s$ and $K_s'$ points of the moir\'e supercell Brillouin zone (BZ), as shown in Fig.~\ref{fig:lattice}(a).
It worth to note that the interlayer distance in TBG varies in real space \cite{uchida-corrugation-prb14}. As schematically shown in Fig.~\ref{fig:lattice}(a), in the $AB (BA)$ region  the interlayer distance $d_{AB}\!\approx\!3.35\,$\angstrom, while  the interlayer distance $d_{AA}\!\approx\!3.6\,$\,angstrom \cite{graphite-AA} in the $AA$-stacked region. Such atomic corrugations can be modeled as \cite{koshino-prx18}
\begin{equation}
d_z(\mathbf{r})=d_{0}+2d_1\sum_{j=1}^{3}\cos{(\,\mathbf{b}_j\!\cdot\!\mathbf{\delta}(\mathbf{r})\,)}\;,
\label{eq:dz-1}
\end{equation}
where $d_0\!=\!3.433\,$\,\angstrom\ and $d_1\!=\!0.0278\,$\angstrom\,
$\mathbf{b}_{1}=(2\pi/a,2\pi/(\sqrt{3}a))$, $\mathbf{b}_2=(-2\pi/a,2\pi/(\sqrt{3}a))$, and $\mathbf{b}_3=\mathbf{b}_1+\mathbf{b}_2$ are the three reciprocal lattice vectors of monolayer graphene. $\mathbf{\delta}(\mathbf{r})$ is a 2D vector indicating the local in-plane shift between the carbon atoms in the two layers around position $\mathbf{r}$ in the moir\'e supercell. In the $AA$ region
$\mathbf{\delta}\!\approx\!(0,0)$ while in the AB region $\mathbf{\delta}\!\approx\!(0,a/\sqrt{3})$.  The corrugations would increase the intersublattice component of the interlayer coupling, and decrease the intrasublattice component of the interlayer coupling.
%We take $d_0\!=\!3.433\,$\,\angstrom\ and $d_1\!=\!0.0278\,$\angstrom\ in order to reproduce the interlayer distances in $AA$- and $AB$-stacked bilayer graphene. S

\section{The non-interacting continuum Hamiltonian}
\label{sec:append2}
The low-energy states of TBG can be  well described by the continuum Hamiltonian proposed in Ref.~\onlinecite{macdonald-pnas11}, 
which is expressed as:
\begin{equation}
H_{\mu}^{0}=\begin{pmatrix}
-\hbar v_F (\mathbf{k}-\mathbf{K}_1^{\mu})\!\cdot\!\mathbf{\sigma}^{\mu} &  U_{\mu}^{\dagger} \\
U_{\mu} &  -\hbar v_F (\mathbf{k}-\mathbf{K}_2^{\mu})\!\cdot\!\mathbf{\sigma}^{\mu}\;,
\end{pmatrix}
\label{eq:h0-mu}
\end{equation} 
where the upper and lower diagonal blocks denote the Dirac fermions in the first and second graphene layers, where $\mathbf{K}_1^{\mu}$ and $\mathbf{K}_2^{\mu}$ are the Dirac points in the first and second graphene layers.
$\mathbf{\sigma}^{\mu}\!=\![\mu\sigma_x, \sigma_y]$, $\mu\!=\!\pm$ is the valley index, with $K^{-}=K$ and $K^{+}=K'$. $U_{\mu}(\mathbf{r})$ denotes the interlayer coupling for the $\mu$ valley:
\begin{equation}
U_{\mu}(\mathbf{r})=\begin{pmatrix}
u_0 g_{\mu}(\mathbf{r}) & u_0'g_{\mu}(\mathbf{r}+\mu\mathbf{r}_{AB})\\
u_0'g_{\mu}(\mathbf{r}-\mu\mathbf{r}_{AB}) & u_0 g_{\mu}(\mathbf{r})
\end{pmatrix} e^{-i\mu\mathbf{\Delta K}\cdot\mathbf{r}}\;,
\label{eq:u}
\end{equation}
where $\mathbf{r}_{AB}\!=\!(\sqrt{3}L_s/3,0)$.  $u_0'$ and $u_0$ denote the intersublattice and intrasublattice interlayer coupling parameters, and $u_0\!<\!u_0'$ if the effects of atomic corrugations are taken into account \cite{koshino-prx18}. In particular, in our calculations $u_0\!=\!0.0797\,$eV, $u_0'\!=\!0.0975\,$eV.
The phase factor $g(\mathbf{r})$ is defined as
$g_{\mu}(\mathbf{r})=\sum_{j=1}^{3}e^{-i\mu\mathbf{q}_j\cdot\mathbf{r}}$,
with $\mathbf{q}_1=(0,-4\pi/3L_s)$, $\mathbf{q}_2=(-2\pi/\sqrt{3}L_s,-2\pi/3L_s)$, and $\mathbf{q}_3=(2\pi/\sqrt{3}L_s,-2\pi/3L_s)$. $\Delta\mathbf{K}\!=\!\mathbf{K}_2-\mathbf{K}_1\!=\![0,4\pi/(3L_s)]$.
 In addition to the $U_v(1)\times SU(2)\times SU(2)$ symmetry as discussed in the main text, the continuum model of each valley  has the symmetry generators $C_{3z}$, $C_{2z}\mathcal{T}$,  and $C_{2x}$,
where $\mathcal{T}$ is the time-reversal operation for spinless fermions (i.e., complex conjugation). The two valleys can be mapped to each other by  $\mathcal{T}$, $C_{2z}$, or $C_{2y}$ operations.

We also consider the situation that TBG is placed on top of an hBN substrate, which is aligned with the bottom graphene layer. This is actually the device used in Refs.~\onlinecite{sharpe-science-19} and \onlinecite{young-tbg-science19}, in which (quantum) anomalous Hall effect has been observed at 3/4 filling of the conduction flat band around the magic angle. The aligned hBN substrate imposes two effects on the electronic structures of TBG. First, the alignment of the hBN substrate with the bottom graphene layer would impose a staggered sublattice potential on the bottom layer graphene and break the $C_{2z}$ symmetry, which  opens a gap at the Dirac points of the flat bands of TBG. Actually the two flat bands per spin for the $K$ valley acquires nonzero Chern numbers $\pm 1$ ($\mp 1$ for the $K'$ valley) once a gap is opened up at the Dirac points. Second, the hBN substrate would generate a new moir\'e pattern, which approximately has the same period as the one generated by the twist of the two graphene layers, but are orthogonal to each other \cite{moon-hbn-graphene-prb14}. However, the moir\'e potential generated by the hBN substrate is one order of magnitude weaker than that generated by the twist of the two graphene layers \cite{moon-hbn-graphene-prb14, jung-hbn-graphene-prb14}, which only generate some additional sub-bands below and above the flat bands of TBG \cite{sharpe-science-19,young-tbg-science19}.
Therefore,  it is a good approximation to neglect the moir\'e potential generated by the hBN substrate. This is actually the approximation widely adopted in previous theoretical studies \cite{zhang-senthil-tbg19, zaletel-tbg-2019}. 
With such an approximation, the effective Hamiltonian for the hBN-aligned TBG system is simplified as
\begin{equation}
H_{\mu}=H^{0}_{\mu}+H_{mass}
\label{eq:h-mu}
\end{equation}
where $H_{mass}$ is the ``Dirac mass" term at the bottom layer graphene generated by the hBN substrate, which is expressed as
\begin{equation}
H_{mass}=\begin{pmatrix}
\Delta\,\sigma_z & 0 \\
0 & 0 
\end{pmatrix}\;.
\label{eq:hmass}
\end{equation}
$\Delta\!=\!15\,$meV is the staggered sublattice potential exerted on the bottom graphene layer. The $H_{mass}$ term would break the $C_{2z}\mathcal{T}$ symmetry associated with each valley, which we will show is crucial in generating the anomalous Hall effect and orbital ferromagnetism.

%The continuum Hamiltonian of TBG for each valley each spin given by Eq.~(\ref{eq:h0-mu}) have $C_{2z}\mathcal{T}$, $C_{3z}$ and $C_{2x}$ symmetries. The $K$ and $K'$ valleys are connected to each other by $\mathcal{T}$, $C_{2z}$, and $C_{2y}$ symmetries. As discussed in the main text, in addition to the crystalline symmetries, there are also separate charge conservation and spin rotational symmetries for each valley, denoted as $U_{v}(1)\times SU(2)\times SU(2)$.

\section{The moir\'e Hartree-Fock formalism}
\label{sec:append3}
The Coulomb interactions between the electrons in TBG can be expressed in momentum space as
\begin{align}
H_{C}=\frac{1}{2N_s}\sum_{\alpha\alpha'}\sum_{\k\k'\q}\sum_{\sigma\sigma'} V(\q)\, \hat{c}_{\k+\q \alpha\sigma}^{\dagger}\hat{c}_{\k'-\q \alpha'\sigma'}^{\dagger} \hat{c}_{\k'\alpha'\sigma'} \hat{c}_{\k\alpha\sigma}
\label{eq:coulomb}
\end{align}
where $\k$ is the atomic wavevector expanded around  the Dirac point in monolayer graphene,  which can be written as $\k=\widetilde{\k}+\mathbf{G}$, where $\widetilde{\k}$ denotes the wavevector with the moir\'e Brillouin zone and $\mathbf{G}$ is a moir\'e reciprocal lattice vector. $\alpha$ is the combined layer and sublattice index, and $\sigma$ is the spin index.  $c^{\dagger}_{\k\alpha\sigma}$ and $c_{\k\alpha\sigma}$ represent the creation and annihilation operators of the Dirac fermions. $N_s$ is the total number of moir\'e supercells in the entire system. $V(\q)$ is Fourier transform of the Coulomb interaction
\begin{align}
V(\q)= \frac{1}{\Omega_{M}}\int d\rr\,\frac{e^2\,e^{-\kappa\vert\rr\vert}}{4\pi\epsilon\epsilon_0 \vert\mathbf{\rr}\vert}e^{-i\q\cdot\rr}\;,
\label{eq:vq}
\end{align}
where $\Omega_M$ is the area of the moir\'e supercell, $\kappa$ is introduced as the inverse screening length, and $\epsilon$ is the  background dielectric constant. $\kappa$ and $\epsilon$ will be treated as two free parameters in the calculations presented in the main text.

If we are interested in the low-energy states around the Dirac points $K$ and $K'$ in graphene, the Dirac fermions can be assigned with the valley index $\mu\!=\!\pm1$. One can re-define the wavevectors $\k$ and $\k'$ as those expanded around the Dirac point $\mathbf{K}^{\mu}$, then the  Coulomb interactions in Eq.~(\ref{eq:coulomb}) can be divided into the intra-valley term $H_{C}^{intra}$ and the inter-valley term $H_{C}^{inter}$,
\begin{align}
H_{C}^{intra}=&\frac{1}{2N_s}\sum_{\alpha\alpha'}\sum_{\mu\mu',\sigma\sigma'}\sum_{\k\k'\q}V(\q) \;\nn
&\times c^{\dagger}_{\k+\q,\mu\sigma\alpha} c^{\dagger}_{\k'-\q,\mu'\sigma'\alpha'}c_{\k',\mu'\sigma'\alpha'}c_{\k,\mu\sigma\alpha}
\label{eq:coulomb-intra}
\end{align}
and 
\begin{align}
H_{C}^{inter}=&\frac{1}{2N_s}\sum_{\alpha\alpha'}\sum_{\mu,\sigma\sigma'}\sum_{\k\k'\q}V(\vert\mathbf{K}-\mathbf{K}'\vert) \;\nn
&\times c^{\dagger}_{\k+\q,\mu\sigma\alpha} c^{\dagger}_{\k'-\q,-\mu\sigma'\alpha'}c_{\k'\mu\sigma'\alpha'}c_{\k,-\mu\sigma\alpha}
\label{eq:coulomb-inter}
\end{align}
Eq.~(\ref{eq:coulomb-intra}) represents the Coulomb scatterings of two electrons which are created and annihilated in the same valley, while Eq.~(\ref{eq:coulomb-intra}) represents some kind of ``pair-hooping" process in momentum space in which two electrons are created in $\mu$ and $-\mu$ and get annihilated in $-\mu$ and $\mu$ valleys. The characteristic interaction strength of $H_{C}^{intra}$ is $V(q=\vert\mathbf{g}_1\vert)\approx 100/\epsilon\,$meV, while the characteristic interaction strength of the intervalley term is $V(\vert\mathbf{K}-\mathbf{K}'\vert)\approx 1.7/\epsilon\,$meV at the magic angle, where $\epsilon$ is the dielectric constant. We thus expect the intravalley interaction would dominate over the intervalley term. The intravalley term $H_{C}^{intra}$ preserves the valley charge conservation and the separate spin rotational symmetry of each valley, i.e., the $U_{v}(1)\times SU(2)\times SU(2)$ symmetry; but the $H_{C}^{inter}$ term only preserves the valley charge conservation and a global $SU(2)$ symmetry, i.e., the $U_{v}(1)\times SU_g(2)$ symmetry.

Now we make Hartree-Fock approximations to Eq.~(\ref{eq:coulomb-intra}) and Eq.~(\ref{eq:coulomb-inter}). The Hartree term of $H^{intra}_C$ reads
\begin{equation}
H_{H}^{intra}=\frac{1}{N_s}\sum_{\mu\sigma}\sum_{\k\alpha}\sum_{\mathbf{Q}}
\,O^{H}(\mathbf{Q})\,c^{\dagger}_{\k+\mathbf{Q},\mu\sigma\alpha} c_{\k,\mu\sigma\alpha}\,
\label{eq:hartree}
\end{equation}
where  the Hartree order parameter $O^{H}(\mathbf{Q})$ associated with the moir\'e reciprocal lattice vector $\mathbf{Q}$ is:
\begin{equation}
O^{H}(\mathbf{Q})=\sum_{\k'\mu'\sigma'\alpha'} V(\mathbf{Q})\langle c^{\dagger}_{\k'-\mathbf{Q},\mu'\sigma'\alpha'} c_{\k',\mu'\sigma'\alpha'}\rangle.
\label{eq:hartree-order}
\end{equation}
The Hartree potential $O^{H}(\mathbf{0})$ represents a spatially homogeneous constant electrostatic potential,  which is expected to be canceled by some positive charge background in the system, thus is dropped in our calculations. Then the leading-order Hartree terms are the six first-neighbor $O^{H}(\mathbf{Q})$ terms with $\vert\mathbf{Q}\vert=4\pi/(\sqrt{3}L_s)$. 

The Fock term of Eq.~(\ref{eq:coulomb-intra}) is expressed as
\begin{align}
H_{F}^{intra}=&-\frac{1}{N_s}\sum_{\mu\sigma\alpha,\mu'\sigma'\alpha'}\sum_{\k,\mathbf{Q}}
O^{F,intra}_{\mu'\sigma'\alpha',\mu\sigma\alpha}(\k,\mathbf{Q}) \;\nn
&\times c^{\dagger}_{\k+\mathbf{Q},\mu\sigma\alpha}c_{\k,\mu'\sigma'\alpha'}
\label{eq:fock}
\end{align}
where the Fock order parameter
\begin{equation}
O^{F,intra}_{\mu'\sigma'\alpha',\mu\sigma\alpha}(\k,\mathbf{Q})
=\sum_{\k'}V(\vert\k+\mathbf{Q}-\k')\langle c^{\dagger}_{\k'-\mathbf{Q},\mu'\sigma'\alpha'} c_{\k',\mu\sigma\alpha}\rangle
\label{eq:fock-order}
\end{equation}

\begin{figure}
\includegraphics[width=3.5in]{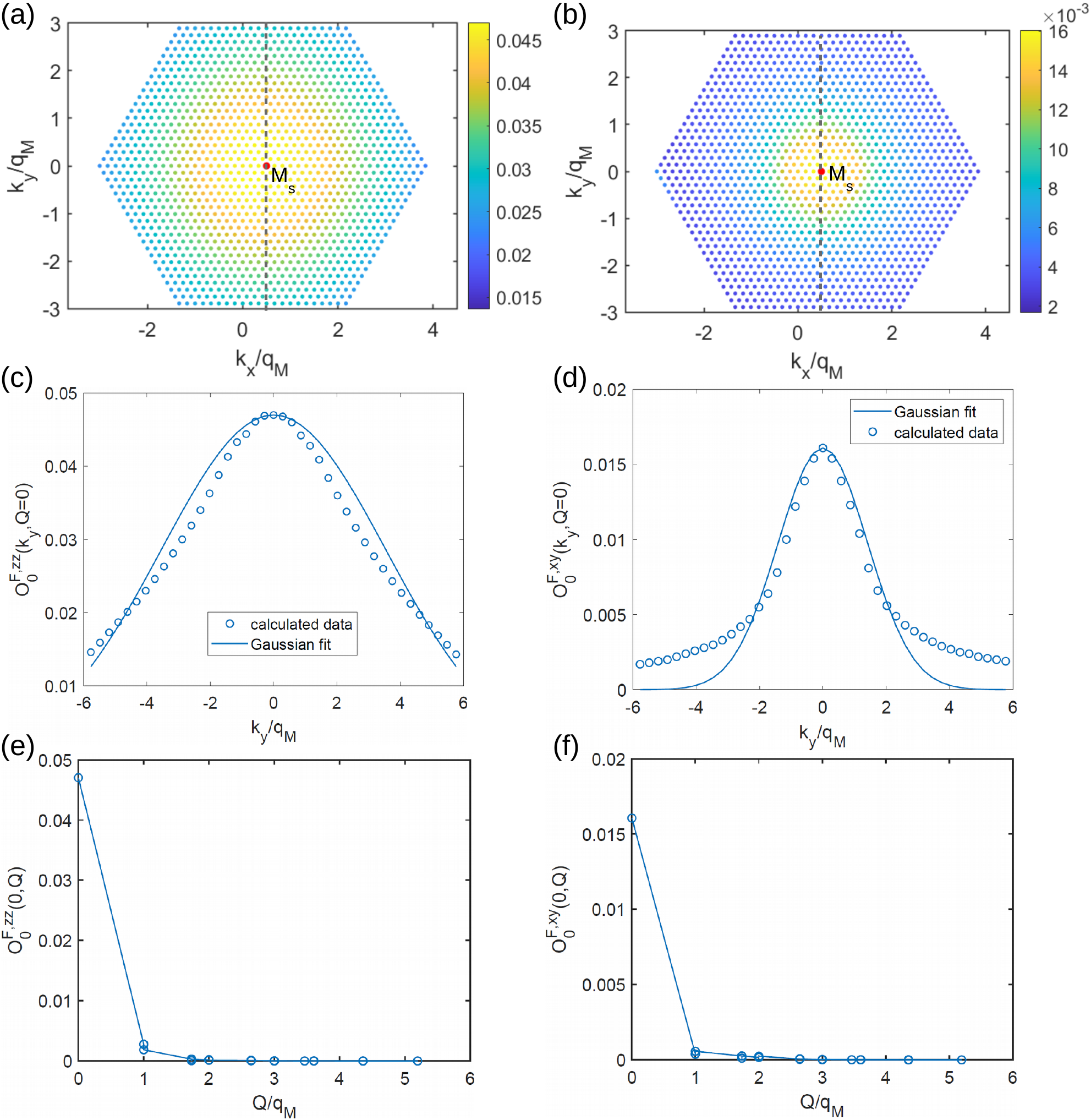}
\caption{The dependence of the Fock order parameters $O^{F,zz}_0(\mathbf{k},\mathbf{Q})$  and  $O^{F,xy}_0(\mathbf{k},\mathbf{Q})$ on atomic wavevector $\k$ and moir\'e reciprocal lattice vector $\mathbf{Q}$. (a) and (b): The $\k$ dependence of the $\vert\mathbf{Q}\vert\!=\!0$ Fock order parameters $O^{F,zz}_0(\mathbf{k},\vert\mathbf{Q}\vert\!=\!0)$ and $O^{F,xy}_0(\mathbf{k},\vert\mathbf{Q}\vert\!=\!0)$ , in units of eV. $k_x$ and $k_y$ are in units of the moire reciprocal lattice constant $q_M=4\pi/(\sqrt{3}L_s)$.  (c) and (d):  $O^{F,zz}_0(\mathbf{k},\vert\mathbf{Q}\vert\!=\!0)$ and $O^{F,xy}_0(\mathbf{k},\vert\mathbf{Q}\vert\!=\!0)$ plotted along the vertical dashed line passing through the $M_s$ point shown in (a) and (b). The blue circles in (c) and (d) represent the actual calculated data, and the blue lines represent Gaussian fittings to the actual data. (e) and (f): The $\vert\mathbf{Q}\vert$ dependence of $O^{F,zz}_0(\mathbf{k},\mathbf{Q})$ and $O^{F,xy}_0(\mathbf{k},\mathbf{Q})$ at $\k=M_s$ ($O^{F,zz}_0(M_s,\mathbf{Q})$ and $O^{F,xy}_0(M_s,\mathbf{Q})$ ), in units of eV.} 
\label{fig:OF}
\end{figure}

Before proceeding, we first check how the Fock self energy depends on the atomic wavevector $\k$ and the moir\'e reciprocal lattice vector $\mathbf{Q}$. In Fig.~\ref{fig:OF}(a)-(c), we show two calculated ``one-step" Fock order parameters:
\begin{align}
O^{F,ij}_0(\mathbf{k},\mathbf{Q})=&\frac{1}{N_M }\sum_{\k'}\sum_{\mu,\mu',s,s', l}V(\vert\k+\mathbf{Q}-\k')\;\nn
&\times\langle c^{\dagger}_{\k'-\mathbf{Q},\mu l s}\tau^i_{\mu\mu'}\sigma^j_{ss'} c_{\k',\mu' l s'}\rangle_0\;,
\label{eq:OF0}
\end{align}
where $c^{\dagger}_{\k'-\mathbf{Q},\mu l s}$ and  $c_{\k',\mu' l s'}$ represent creation and annihilation operators of electrons, with $\mu, l$ and $s$ being the valley, layer, and sublattice indices.  $\langle ...\rangle_0$ stands for the expectation value calculated using the Bloch states of the continuum model with  a constant  ($\k$ independent) Fock self-energy $\Delta_0\tau^{i}\sigma^{j}$  ($\Delta_0$ is chosen as 10\,meV) being applied, and the chemical potential is fixed at zero filling. $\tau^{i}$ and $\sigma^{j}$ are the Pauli matrices in the valley and sublattice space ($i, j=0, x, y, z$).  
In other words, $O^{F,ij}_0(\mathbf{k},\mathbf{Q})$ represents the ``one-step" Fock order parameter (without self consistency), which are calculated exactly without any assumption using Eq.~(\ref{eq:OF0}). 
In Fig.~\ref{fig:OF}(a), (b) we show the magnitudes of  $O^{F,zz}_0(\mathbf{k},\mathbf{Q}\!=\!0)$ and $O^{F,xy}_0(\mathbf{k},\mathbf{Q}\!=\!0)$ in the 2D plane of the atomic wavevector $\k=$ centered at the untwisted graphene Dirac point $K$, which is folded to the moire $M_s$ point as marked by the red dot in Fig.~\ref{fig:OF}(a) and (b). $O^{F,zz}_0(\mathbf{k},\mathbf{Q})$ is a valley-sublattice polarized order, while $O^{F,xy}_0(\mathbf{k},\mathbf{Q})$ is the ``Kramers intervalley coherent order" proposed in Ref.~\onlinecite{zaletel-tbg-hf}.
We see that both of the ``one-step" Fock order parameters  are exponentially localized near the untwisted Dirac point $K$ ($M_s$).  This is because $O^{F,zz}_0(\mathbf{k},\mathbf{Q}\!=\!0)$ and $O^{F,xy}_0(\mathbf{k},\mathbf{Q}\!=\!0)$ are calculated using the Hamiltonians with constant ($\k$ independent) Fock order parameters $\Delta_0\tau^{z}\sigma^{0}$ and $\Delta_0\tau^{x}\sigma^{y}$ ($\Delta_0\!=\!10\,$meV) applied to the non-interacting continuum model, which open gaps at the charge neutrality point. The Fock self energy of such gapped states thus decay exponentially in $\k$ space.  We expect such a behavior is generic for all gapped states, therefore we fit the $\k$ dependence of the Fock order parameter using a Gaussian function as given by Eq.~(4) in the main text.

%In Fig.~\ref{fig:OF}(d)-(e) we take line cuts of 

In Fig.~\ref{fig:OF}(c)-(d), we plot the magnitudes of  $O^{F,zz}_0(\mathbf{k},\mathbf{Q}\!=\!0)$ and $O^{F,xy}_0(\mathbf{k},\mathbf{Q}\!=\!0)$ as a function of $k_y$ at $k_x=q_M/2$ (along the gray dashed line passing through $M_s$ in Fig.~\ref{fig:OF}(a) and (b) ), where $q_M=4\pi/(\sqrt(3)Ls)$. The blue circles in Fig.~\ref{fig:OF}(c)-(d) are the actual data calculated using Eq.~(\ref{eq:OF0}), while the blue lines represent  Gaussian fittings to the calculated data. We see that the Gaussian fitting works well for both type of orders, thus it is legitimate to assume that $O^{F,\mu}_0(\mathbf{k},\mathbf{Q}\!=\!0)$ decays with $\vert\mathbf{\k}\vert$ in a Gaussian form especially for gapped states.  In Fig.~\ref{fig:OF}(e) and (f) we show the $\vert\mathbf{Q}\vert$
dependence of $O^{F,zz}_0(\mathbf{k},\mathbf{Q})$ and $O^{F,xy}_0(\mathbf{k},\mathbf{Q})$  at $\mathbf{k}\!=\!M_s$. It is clearly shown  that magnitudes of $O^{F,zz}_0(M_s, \mathbf{Q})$ and $O^{F,xy}_0(M_s,\mathbf{Q})$ decay rapidly with the increase of $\vert\mathbf{Q}\vert$: it drops from 47\,meV (16\,meV) to 2.7\,meV (0.5\,meV) when $\vert\mathbf{Q}\vert$ is increased from zero to $q_M$ for $O^{F,zz}_0$ ($O^{F,xy}_0$).

It follows from Fig.~\ref{fig:OF}(e)-(f) that the $\vert\mathbf{Q}\vert\neq 0$ Fock terms are negligible compared with the $\vert\mathbf{Q}\vert\!=\!0$ term. Therefore, 
it is an excellent approximation to keep only the $\vert\mathbf{Q}\vert=0$ term of $O^{F,intra}_{\mu'\sigma'\alpha',\mu\sigma\alpha}(\k,\mathbf{Q})$, i.e.,
\begin{align}
H_{F}^{intra}\approx -\frac{1}{N_s}\sum_{\mu\sigma\alpha,\mu'\sigma'\alpha'}
\sum_{\k}& O^{F,intra}_{\mu'\sigma'\alpha', \mu\sigma\alpha}(\k,\mathbf{Q}\!=\!\mathbf{0})\;\nn
&\times c^{\dagger}_{\k,\mu\sigma\alpha} c_{\k,\mu'\sigma'\alpha'}\;.
\label{eq:fock-local}
\end{align}
Moreover, according to the Gaussian-like decay behavior shown in Fig.~\ref{fig:OF}(c)-(d), we further assume that the $\k$ dependence of the Fock order parameter $O^{F,intra}_{\mu'\sigma'\alpha', \mu\sigma\alpha}(\k)$ can be described by a simple Gaussian function as shown in Eq.~(4) in the main text.   As already discussed in the main text, we also assume that the Fock order parameters are independent of the layer index, since  the two layers have already been equally mixed by the moir\'e potential to give rise to the non-interacting flat bands at the magic angle \cite{jpliu-prb19}.  
Therefore, in the end, the intravalley Fock terms are dependent on the valley, spin, and sulattice indices, which have 64 independent order parameters. The spread of the Fock order parameters in $\k$ space is considered as an extra variational parameter.

Next we consider the Fock term of $H_C^{inter}$, which can be expressed as
\begin{equation}
H_{F}^{inter}=-\frac{1}{N_s}\sum_{\k}\sum_{\mu\sigma\sigma'\alpha\alpha'}
\sum_{\k} O^{F,inter}_{\mu\sigma\alpha,\mu\sigma'\alpha'}\,c^{\dagger}_{\k,-\mu\sigma'\alpha'}c_{\k,-\mu\sigma\alpha}\;,
\label{eq:fock-inter}
\end{equation}
where 
\begin{equation}
O^{F,inter}_{\mu\sigma\alpha,\mu\sigma'\alpha'}=\sum_{\k'} V(\vert\mathbf{K}-\mathbf{K'}\vert)\langle c^{\dagger}_{\k',\mu\sigma\alpha} c_{\k',\mu\sigma'\alpha'}\rangle\;,
\label{eq:fock-inter-order}
\end{equation}
Note that, Eq.~(\ref{eq:fock-inter-order}) is vanishing for intervalley coherent (IVC) type of orders which break the valley charge conservation. Therefore, neglecting the Hartree term of the intervalley interaction, only the intra-valley interaction would give rise to the IVC states, which break the valley charge conservation ($U_v(1)$ symmetry) and the relative spin rotational symmetry of the two valleys ($SU_v(2)$ symmetry). %Including the Hartree term of $H_C^{inter}$ would slightly break the degeneracy of the IVC ordered states, but the splitting is tiny, and it barely changes our conclusions.

We define the Hartree-Fock  operator as $V_{HF}=H_{H}^{intra}+H_{F}^{intra}+H_{F}^{inter}$, and the total mean-field Hartree-Fock Hamiltonian $H_{MF}=\sum_{\mu,\sigma} H_{\mu} + V_{HF}$, which is just Eq.~(\ref{eq:hmf}) in the main text. In particular, in Eq.~(\ref{eq:hmf}), $H_0=\sum_{\mu,\sigma}H_{\mu}$ is the non-interacting Hamiltonian for TBG, $H_{\mu}$ is just the spin-independent continuum Hamiltonian for valley $\mu$ given by Eq.~(\ref{eq:h0-mu})-(\ref{eq:h-mu}); the Fock operator $O^{F}$ in Eq.~(\ref{eq:hmf}) includes both $O^{F,intra}$ and $O^{F,inter}$.  Then we determine the ground-state wavefunction $\vert\psi\rangle$ using the mean-field Hamiltonian of Eq.~(\ref{eq:hmf}), and 
treat the Hartree-Fock order parameters $\mathcal{O}^{HF}=\{O^{H}(\mathbf{Q}), O^{F}_{\mu\sigma\alpha,\mu'\sigma'\alpha'}(\k)\}$ discussed above as variational parameters, which are determined by minimizing the total energy functional:
\begin{equation}
E_T(\mathcal{O}^{HF})=\langle \psi(\mathcal{O}^{HF})\vert H_0 + H_C \vert \psi(\mathcal{O}^{HF})\rangle\;.
\end{equation}
Again, $\vert \psi(\mathcal{O}^{HF})\rangle$ is the ground state of the effective mean-field Hamiltonian $H_{MF}$, which are dependent on the variational Hartree-Fock order parameters $\mathcal{O}^{HF}=\{O^{H}(\mathbf{Q}), O^{F}_{\mu\sigma\alpha,\mu'\sigma'\alpha'}(\k)\}$. $H_C$ is the physical Coulomb interaction term given in Eq.~(\ref{eq:coulomb}). It is worthwhile to note that  $\mathbf{k}$ (or $\k'$) in the above equations are wavevectors defined in the monolayer graphene BZ, which can be re-written as $\k\!=\!\kt+\mathbf{G}$, where $\kt$ is the wavevector in the moir\'e BZ,  $\mathbf{G}$ is  a moir\'e reciprocal lattice vector. Then the above formalism can be implemented in the plane-wave basis. e

To be specific, for each integer filling and each $(\epsilon, \kappa)$ parameter, we perform two self-consistent calculations starting from different initial wavefunctions: one starts from a generic valley-polarized state including all the $\tau_0 s_j \sigma_k$ and $\tau_z s_j \sigma_k$  order parameters ($j, k=0, x, y, z$),  and the other starts from a generic intervalley-coherent state including all the $\tau_x s_j \sigma_k$ and $\tau_y s_j \sigma_k$  order parameters. We compare the energies starting from the two different categories of trial states, and determine the ground state. The minimization of energy with respect to multiple variational parameters is performed  by the \textit{optimize.minimize} function in SciPy using quasi-Newton method. Partial derivatives of the total energy with respect to the variational parameters are expressed in terms of the eigenergies and density matrices at each HF iteration, and are provided to the \textit{optimize.minimize} function. This guarantees that the iteration process flows along the descending direction of the total energy,  and that one would reach a genuine energy minimum at convergence.    The massive calculations are performed on a $3\times 3$ $\widetilde{\mathbf{k}}$ mesh in the moir\'e Brillouin zone. We have re-performed calculations for a few representative ($\epsilon$, $\kappa$) points at each integer filing on a $6\times 6$ $\widetilde{\mathbf{k}}$ mesh, and find that the conclusions are unchanged. 
The convergence threshold for the total energy is set to $10^{-7}\,$eV.

\bibliography{tmg}

\widetext
\clearpage
\begin{center}
\textbf{\large Supplemental Material for ``Theories for the correlated insulating states and quantum anomalous Hall effect in twisted bilayer graphene"}
\end{center}

\vspace{12pt}
\begin{center}
\textbf{\large \I\ The Hartree-Fock bandstructures at 1/2 filling of TBG}
\end{center}

\begin{figure}
\includegraphics[width=3.5in]{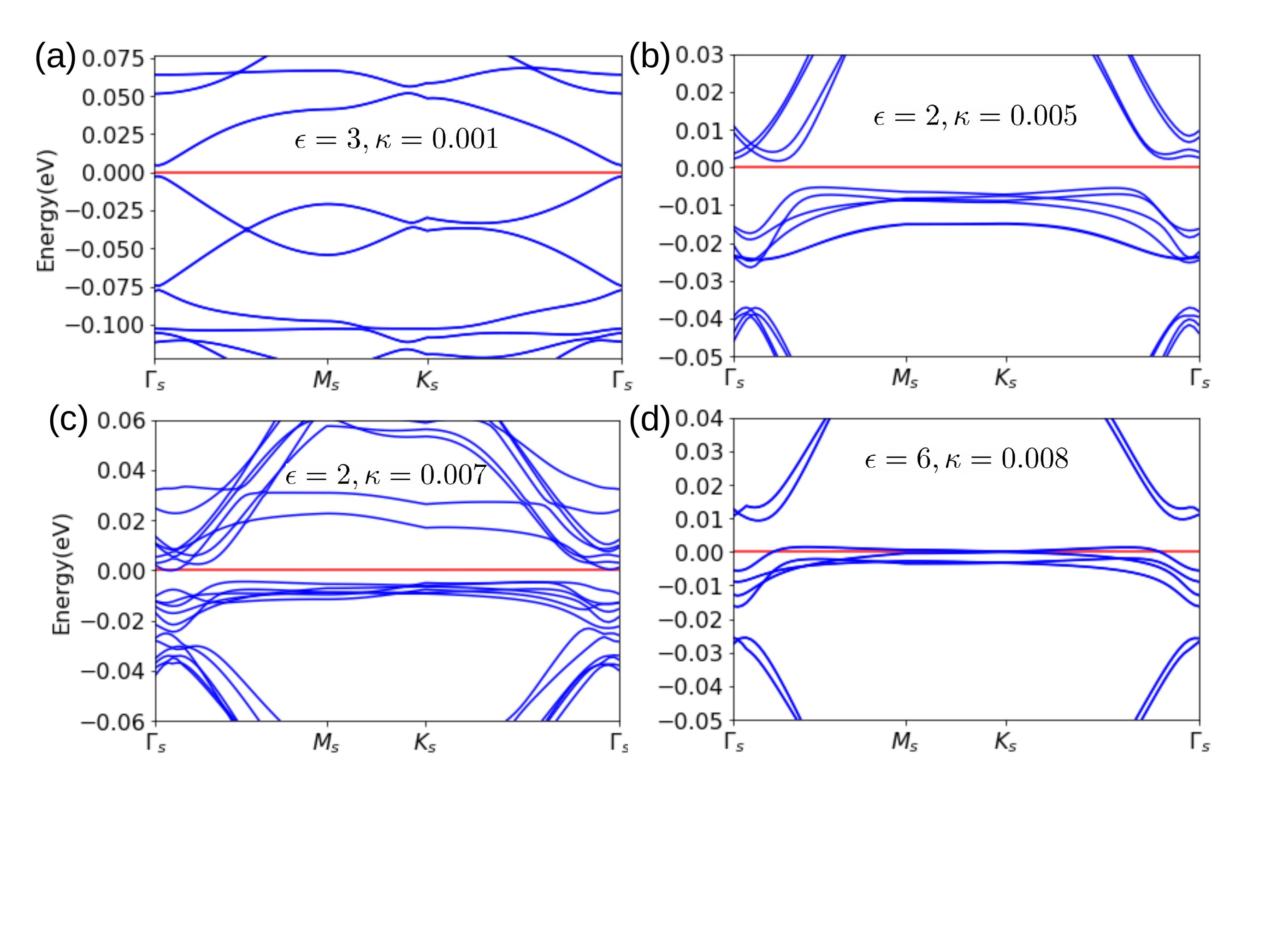}
\caption{Band structures for the Hartree-Fock ground states of TBG at 1/2 filling at the magic angle: (a) $\epsilon\!=\!3, \kappa\!=\!0.001\,\angstrom^{-1}$, (b) $\epsilon\!=\!2, \kappa\!=\!0.005\,\angstrom^{-1}$,  (c) $\epsilon\!=\!3, \kappa\!=\!0.007\,\angstrom^{-1}$, and (d) $\epsilon\!=\!6, \kappa\!=\!0.008\,\angstrom^{-1}$, where $\epsilon$ is the dielectric constant, and $\kappa$ is the inverse screening length.}
\label{fig:filling6-bands}
\end{figure}

In Fig.~\ref{fig:filling6-bands} we show the bandstructures of the Hartree-Fock ground states (with local-exchange approximation) for the magic-angle twist bilayer graphene (TBG) at 1/2 filling in the $\delta$, $\epsilon$, $\gamma$, and $\alpha$ phases (see Fig.~2 and Table~\I\ of the main text for the definitions of these phases). 
The bandstructures for the correlated insulating states in the $\beta$ phase have already been discussed in detail in the main text.  For $\kappa\!=\!0.001\,\angstrom^{-1}$ and $\epsilon\!=\!3$, the system is in the $\delta$ phase with an indirect gap $\sim\!0.0085\,$eV, and the dominant order parameters are the intervalley coherent (IVC) orders $\Delta^{xjx}\,\tau_x\,s_j\,\sigma_x$ and $\Delta^{yjx}\tau_y\,s_j\,\sigma_x$ ($j\!=\!x,y,z$) with $\Delta^{xjx}, \Delta^{yjx}\!\sim$70\,meV, and the corresponding bandstructures are given in Fig.~\ref{fig:filling6-bands}(a). Such an order parameter breaks the valley charge conservation ($U_v(1)$ symmetry), and the relative spin rotational symmetry of the two valleys ($SU_v(2)$ symmetry), and it preserves the global $SU(2)$ symmetry, $\mathcal{T}$ symmetry, and all the crystalline symmetries. The global $SU(2)$ symmetry of such an ordered state can be seen as follows:
the full Hartree-Fock (HF) Hamiltonian with the generic IVC order parameters $\tau_x\s_j$ and/or $\tau_y\s_j$ ($j=0,x,y,z$) can always be transformed to the one with the spin-independent order parameter $\tau_xs_0$ and/or $\tau_y s_0$ by two independent spin rotations (represented by $U_{K}$ and $U_K'$) in the two valleys: 
\begin{equation}
\tau_x s_0 =\begin{pmatrix}
0 & s_0 \;\\
s_0 & 0 
\end{pmatrix}=\begin{pmatrix}
U_K^{\dagger} & 0\;\\
0 & U_{K'}^{\dagger} 
\end{pmatrix} \,\times\,\begin{pmatrix}
0 & s_j\;\\
s_j & 0
\end{pmatrix}\, 
\times\,\begin{pmatrix}
U_K & 0\;\\
0 & U_{K'}
\end{pmatrix}\;,
\label{eq:unitary-transform}
\end{equation}
i.e., one can always find two $2\times 2$ unitary matrices $U_K$ and $U_K'$ such that $s_0\!=\!U_K^{\dagger}s_j U_{K'}\!=\!U_{K'}^{\dagger}s_jU_{K}$ with $j=x, y, z$. Physically, it means one can always make the intervalley coupling being spin independent by redefining the spin quantization axes of the two valleys.
Every band of the HF Hamiltonian with $\tau_{x(y)}s_0$-type of order parameters is twofold spin degenerate, and such twofold degeneracy cannot be lift by unitary transformations shown in Eq.~(\ref{eq:unitary-transform}). Therefore, the band structures of the HF Hamiltonians with the generic IVC order parameters would remain spin degenerate, as clearly shown in Fig.~\ref{fig:filling6-bands}(a). 

For $\kappa\!=\!0.005$ and $\epsilon\!=\!2$, at 1/2 filling, the system is in the $\epsilon$ phase (see Fig.~3 and Table~\I\ of the main text), and the dominant order parameters are the VP-type orders with $\tau_0\,\mathbf{s}_{\hat{\mathbf{n}}}\,\sigma_z$  and $\tau_z\,\mathbf{s}_{\hat{\mathbf{n}}}\,\sigma_z$, where $\mathbf{s}_{\hat{\mathbf{n}}}$ is the Pauli matrix representing the spin operator pointing along the direction $\hat{\mathbf{n}}$. The choice $\hat{\mathbf{n}}$ is arbitrary due to the $SU(2)$ degeneracy.  The corresponding bandstructures are shown in Fig.~\ref{fig:filling6-bands}(b).
Such ordered states are valley polarized states with antiferromagnetic-like spin ordering on the $A$ and $B$ sublattices, which break $\mathcal{T}$ symmetry and spin rotational symmetry.

Sometimes the VP order and the IVC order may co-exist in the phase diagram of Fig.~3 in the main text. For example, when $\epsilon\!=\!2$, and $\kappa\!=\!0.007\,\angstrom^{-1}$, the system is in the $\gamma$ phase (see Fig.~3 and Table~\I\ of the main text), and the ground state is insulating with a gap $\sim 7.5$\,meV. The order parameters include both the IVC and VP types of orders: $\Delta_{IVC}\,\sum_{j=x,y,z}(\tau_x\,s_j\,\sigma_x + \tau_y\,s_j\,\sigma_x) + \Delta_{VP}\,\tau_z\,s_0\,\sigma_x$, where both $\Delta_{IVC}$ and  $\Delta_{VP}$ are around 10\,meV. Such a state  with co-existing IVC and VC orders breaks $\mathcal{T}$ and spin rotational symmetry, and it does not have any special symmetry protecting the band degeneracy. Thus each band at a generic $\mathbf{k}$ point is non-degenerate as shown in Fig.~\ref{fig:filling6-bands}(c). However, the  state with co-existing IVC and VP orders still has an \textit{orbital} $C_{2z}\mathcal{T}$ symmetry represented by $\tau_z \,s_0\,\sigma_z\,\mathcal{K}$.  The orbital $C_{2z}\mathcal{T}$ symmetry enforces that the AHC and orbital magnetization must vanish, such that the system is  a $C\!=\!0$ trivial insulator.

For $\epsilon\!=\!6$ and $\kappa\!=\!0.008\,\angstrom^{-1}$, at 1/2 filling the system is  in the $\alpha$ phase as shown in Fig.~3 of the main text.  The dominant order parameter is $\Delta^{z0x}\,\tau_z\,s_0\,\sigma_x$ with $\Delta^{z0x}\!\sim\!-5\,$meV. Such a valley-polarized ordered state breaks both $\mathcal{T}$ and $C_{2z}$ symmetry due to the valley polarization, but preserves the combined $C_{2z}\mathcal{T}$ symmetry, which kills the anomalous Hall effect.
The band structures of the valley-splitted metallic phase are shown in Fig.~\ref{fig:filling6-bands}(d). We see that every band is spin degenerate, and the system remains metallic.

\hspace{24pt}
%\section{Hartree-Fock bandstructures of hBN-aligned TBG at different fillings}
\begin{center}
\textbf{\large \II\ Hartree-Fock bandstructures of hBN-aligned TBG at different fillings}
\end{center}

\begin{figure}
\includegraphics[width=3.5in]{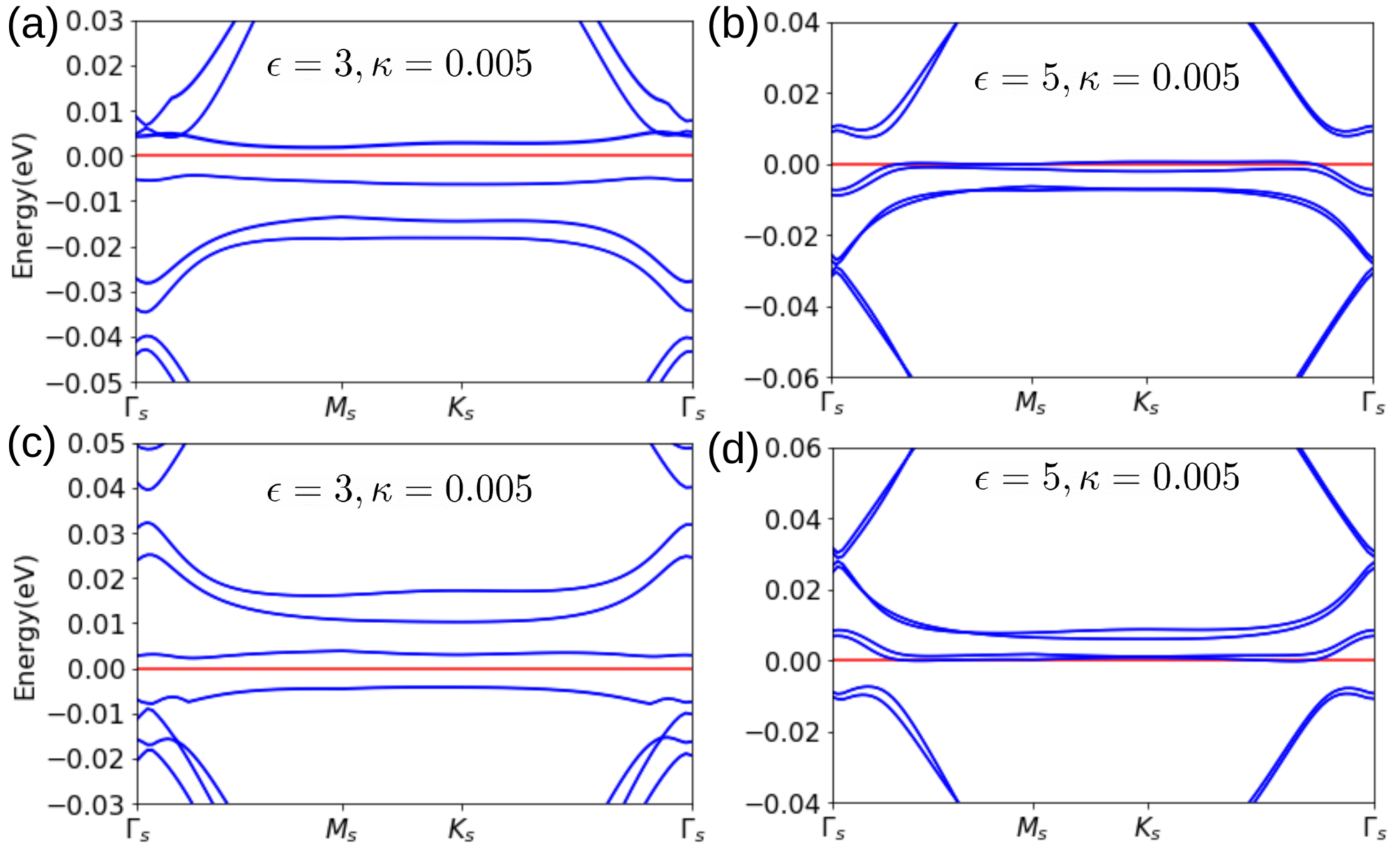}
\caption{Band structures for the Hartree-Fock ground states of TBG aligned with hBN substrate at the magic angle: (a) at 1/2 filling, $\epsilon\!=\!3, \kappa\!=\!0.005\,\angstrom^{-1}$, (b) at 1/2 filling, $\epsilon\!=\!5, \kappa\!=\!0.005\,\angstrom^{-1}$,  (c) at -1/2 filling, $\epsilon\!=\!3, \kappa\!=\!0.005\,\angstrom^{-1}$, and (d) at -1/2 filling, $\epsilon\!=\!5, \kappa\!=\!0.005\,\angstrom^{-1}$, where $\epsilon$ is the dielectric constant, and $\kappa$ is the inverse screening length.}
\label{fig:filling26-bands}
\end{figure}

In Fig.~\ref{fig:filling26-bands} we show the Hartree-Fock bandstructures (with local-exchange approximation) of hBN-aligned substrate at $\pm 1/2$ fillings. To be specific, in Fig.~\ref{fig:filling26-bands}(a) and (c) we show the bandstructures at 1/2 and -1/2 fillings with $\epsilon\!=\!3$, $\kappa\!=\!0.005\,\angstrom^{-1}$, which are the quantum anomalous Hall insulating states with Chern number  $-2$ and $2$ respectively. In Fig.~\ref{fig:filling26-bands}(b) and (d) we show the bandstructures at $1/2$ and $-1/2$ fillings with $\epsilon\!=\!5$, $\kappa\!=\!0.005^{-1}$. As the interaction becomes weaker for larger dielectric constant and larger inverse screening length, the system becomes metallic at $\pm 1/2$ filling, but still exhibits non-vanishing anomalous Hall effect due to the breaking of $C_{2z}\mathcal{T}$ symmetry. 

\begin{figure}
\includegraphics[width=3.5in]{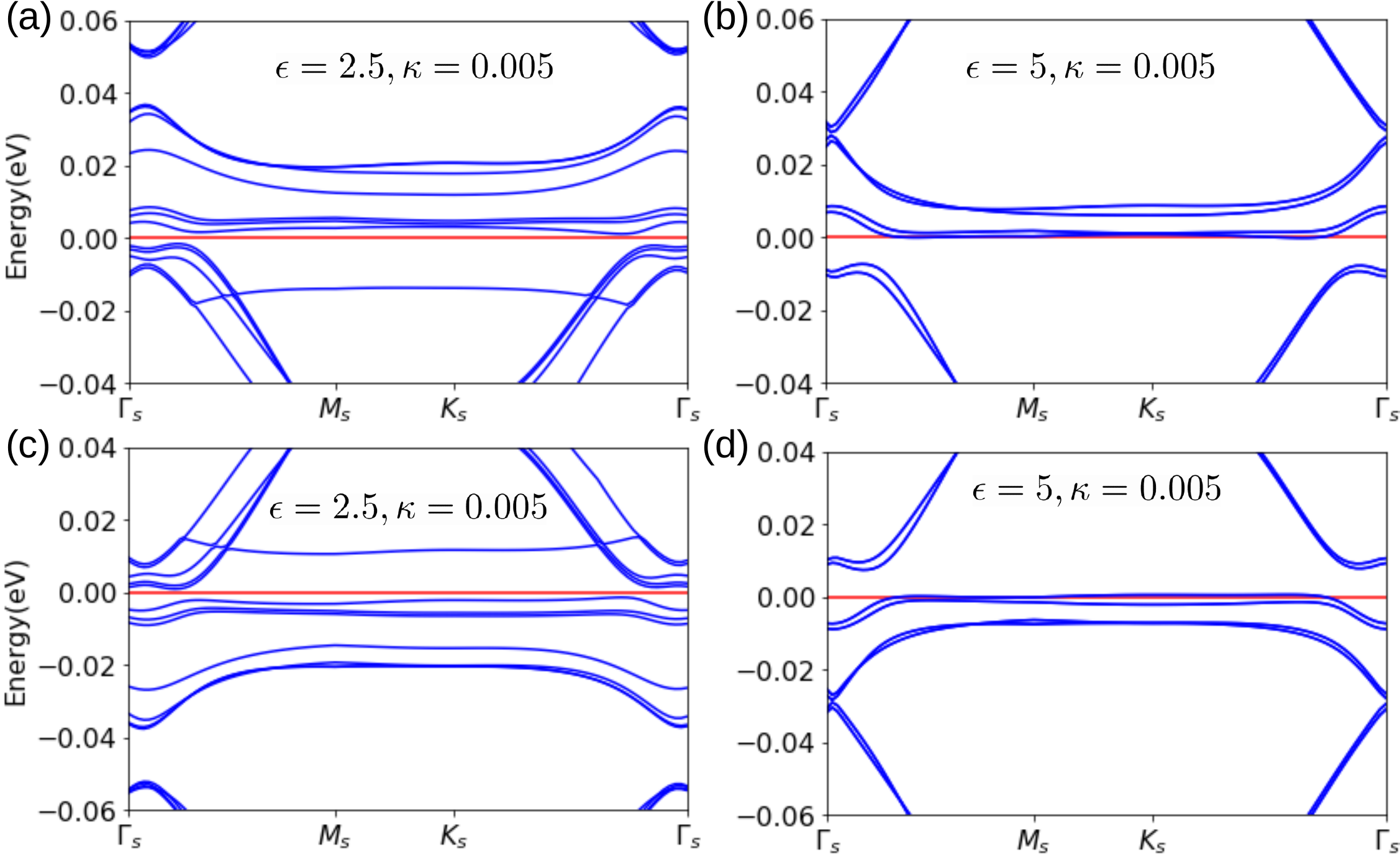}
\caption{Band structures for the Hartree-Fock ground states of TBG aligned with hBN substrate at the magic angle: (a) at -$3/4$ filling, $\epsilon\!=\!2.5, \kappa\!=\!0.005\,\angstrom^{-1}$, (b) at -3/4 filling, $\epsilon\!=\!5, \kappa\!=\!0.005\,\angstrom^{-1}$,  (c) at 3/4 filling, $\epsilon\!=\!2.5, \kappa\!=\!0.005\,\angstrom^{-1}$, and (d) at 3/4 filling, $\epsilon\!=\!5, \kappa\!=\!0.005\,\angstrom^{-1}$, where $\epsilon$ is the dielectric constant, and $\kappa$ is the inverse screening length.}
\label{fig:filling17-bands}
\end{figure}

In Fig.~\ref{fig:filling17-bands} we show the bandstructures of hBN-aligned TBG at $\pm 3/4$ fillings. In Fig.~\ref{fig:filling17-bands}(a) and (c) we show the bandstructures at -3/4 and 3/4 fillings with $\epsilon\!=\!2.5$, $\kappa\!=\!0.005\,\angstrom^{-1}$, which are the quantum anomalous Hall insulating states with Chern number $+1$ and $-1$ respectively. In Fig.~\ref{fig:filling17-bands}(b) and (d) we show the bandstructures at $-3/4$ and $3/4$ fillings with $\epsilon\!=\!5$, $\kappa\!=\!0.005\,\angstrom^{-1}$, at which the system becomes metallic due to the reduced interaction strength.

\vspace{12pt}
\begin{center}
\textbf{\large \III\ The twist-angle dependence of the correlated insulating states and the quantum anomalous Hall effects}
\end{center}

\begin{figure}
\includegraphics[width=2.5in]{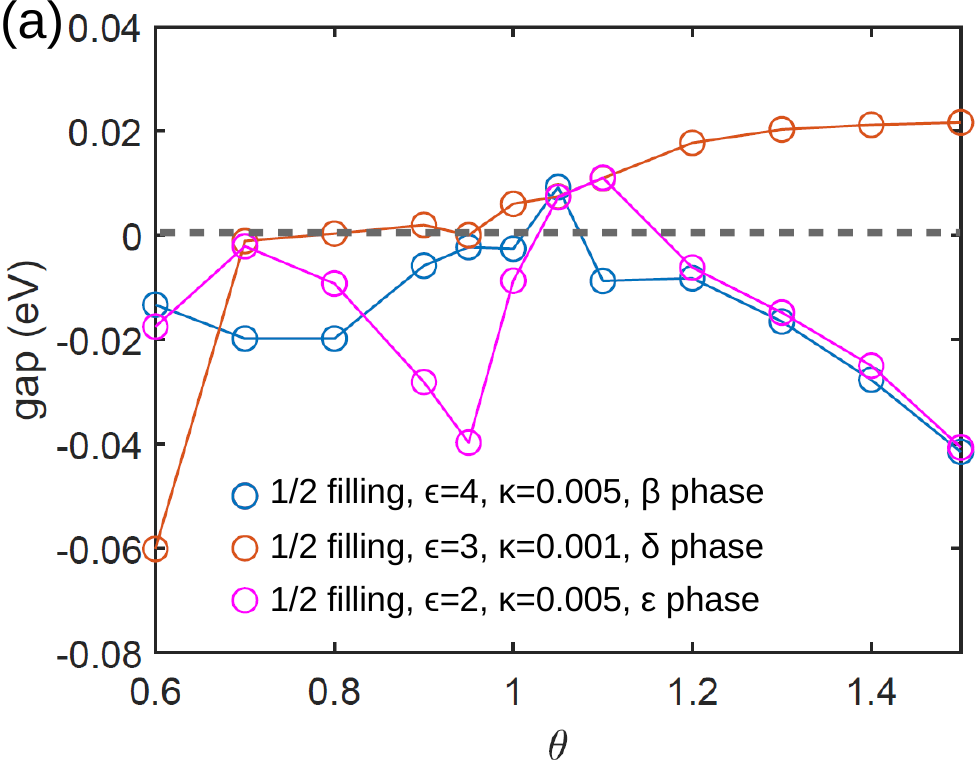}
\caption{Twist-angle dependence of the indirect gaps of the correlated insulating states at 1/2 filling of TBG. The blue, red, and magenta circles denote the cases for $\epsilon\!=\!4, \kappa\!=\!0.005\,\angstrom^{-1}$, $\epsilon\!=\!3, \kappa\!=\!0.001\,\angstrom^{-1}$, and $\epsilon\!=\!2, \kappa\!=\!0.005\,\angstrom^{-1}$, respectively.}
\label{fig:gap-varyangle}
\end{figure}

In this section we study the twist-angle dependence of the correlated insulating states and the quantum anomalous Hall effects. In Fig.~\ref{fig:gap-varyangle} we show  the dependence of the indirect gaps of the Hartree-Fock ground states at 1/2 filling of TBG  with respect to  the twist angle ($\theta$), where the blue, red, and magenta circles represent the cases for the interaction parameters  $\epsilon\!=\!4$, $\kappa\!=\!0.005\,\angstrom^{-1}$,  $\epsilon\!=\!3$, $\kappa\!=\!0.001\,\angstrom^{-1}$,  and $\epsilon\!=\!2$, $\kappa\!=\!0.005\,\angstrom^{-1}$ respectively. When $\epsilon\!=\!4$, $\kappa\!=0.005\,\angstrom^{-1}$,  the system in the $\beta$  phase, we see that the gap induced by the valley polarization in the $\beta$ phase is nonzero \textit{only at the magic angle}. This is because the bandwidth is minimal at the magic angle, which significantly enhances the interaction effects.  Such twist-angle dependence is also consistent with the experimental observation that the correlated insulating state appears only at the magic angle. When $\epsilon\!=\!3$, $\kappa\!=\!0.01\,\angstrom^{-1}$, the  system is in the  $\delta$ phase with intervalley coherent order.  In such a state, the gap generically exists for twist angle $\theta\gtrapprox 1^{\circ}$, and this is inconsistent experimental observation. When $\epsilon\!=\!2$, $\kappa\!=\!0.005\,\angstrom^{-1}$, the system is in the $\varepsilon$ phase with spin antiferromagnetic ordering on $A$ and $B$ sublattices. The gap in such a state also only occurs at the magic angle, but the insulating state in the $\varepsilon$ phase only occurs in a very small region of the parameter space as shown in Fig.~3 of the main text, and the gap in such a spin antiferromagnetic state does not decrease with the increase of external magnetic field (data not shown), which is inconsistent with experimental observations.
It follows that the  the experimentally observed correlated insulating state is most likely to be the valley polarized insulating state in the $\beta$ phase.

\begin{figure}
\includegraphics[width=2.8in]{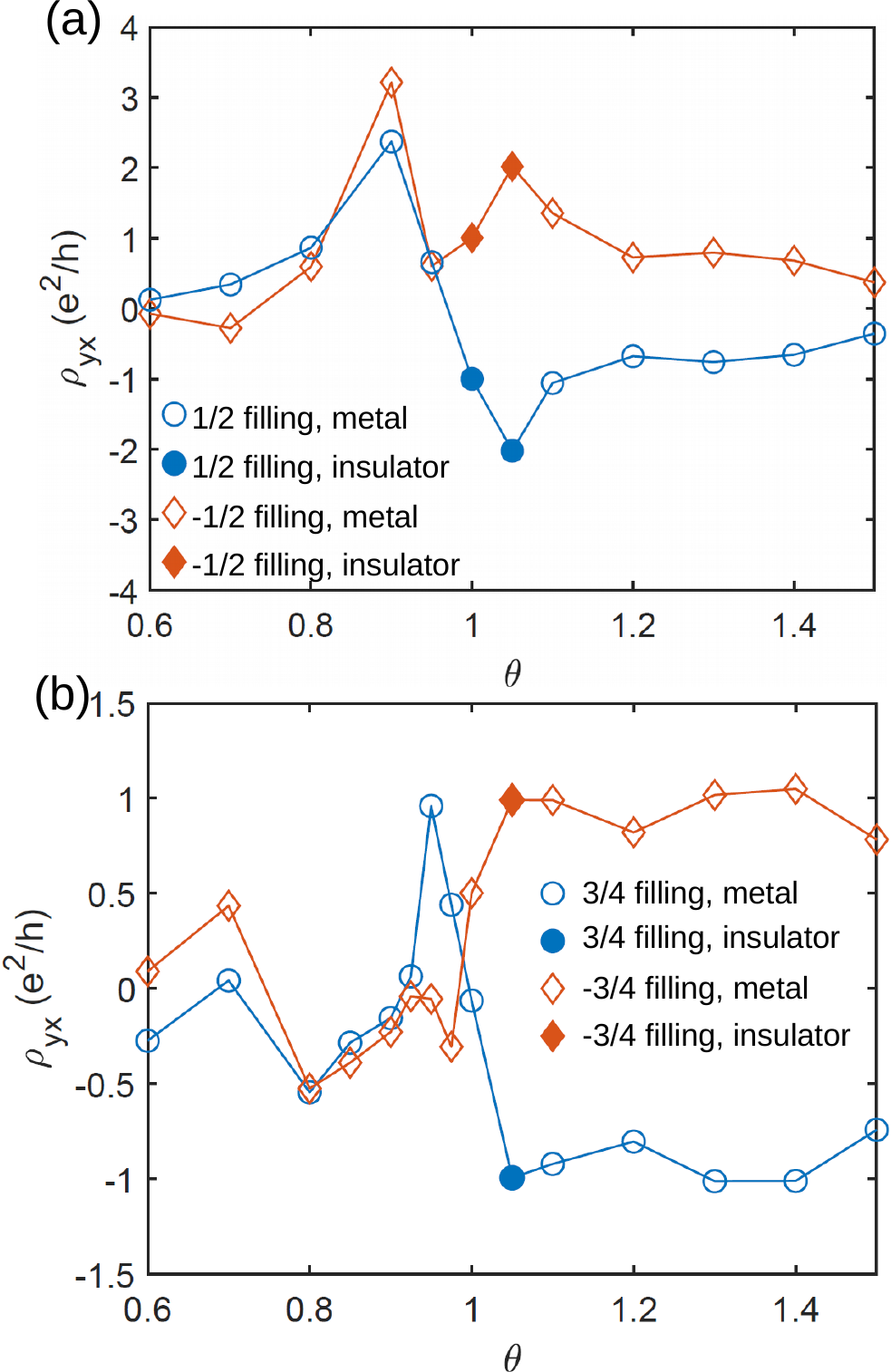}
\caption{The twist-angle dependence of the anomalous Hall conductivities of the Hartree-Fock ground states of hBN-aligned TBG: (a) at $\pm1/2$ fillings, and (b) at $\pm 3/4$ fillings. The blue circles and red diamonds in (a) denote the $1/2$ and $-1/2$ fillings , while in (b) they denote the $3/4$ and $-3/4$ fillings respectively. The filled (open) symbols indicate the states are insulating (metallic). }
\label{fig:ahc-varyangle}
\end{figure}
\begin{figure*}
\includegraphics[width=5.0in]{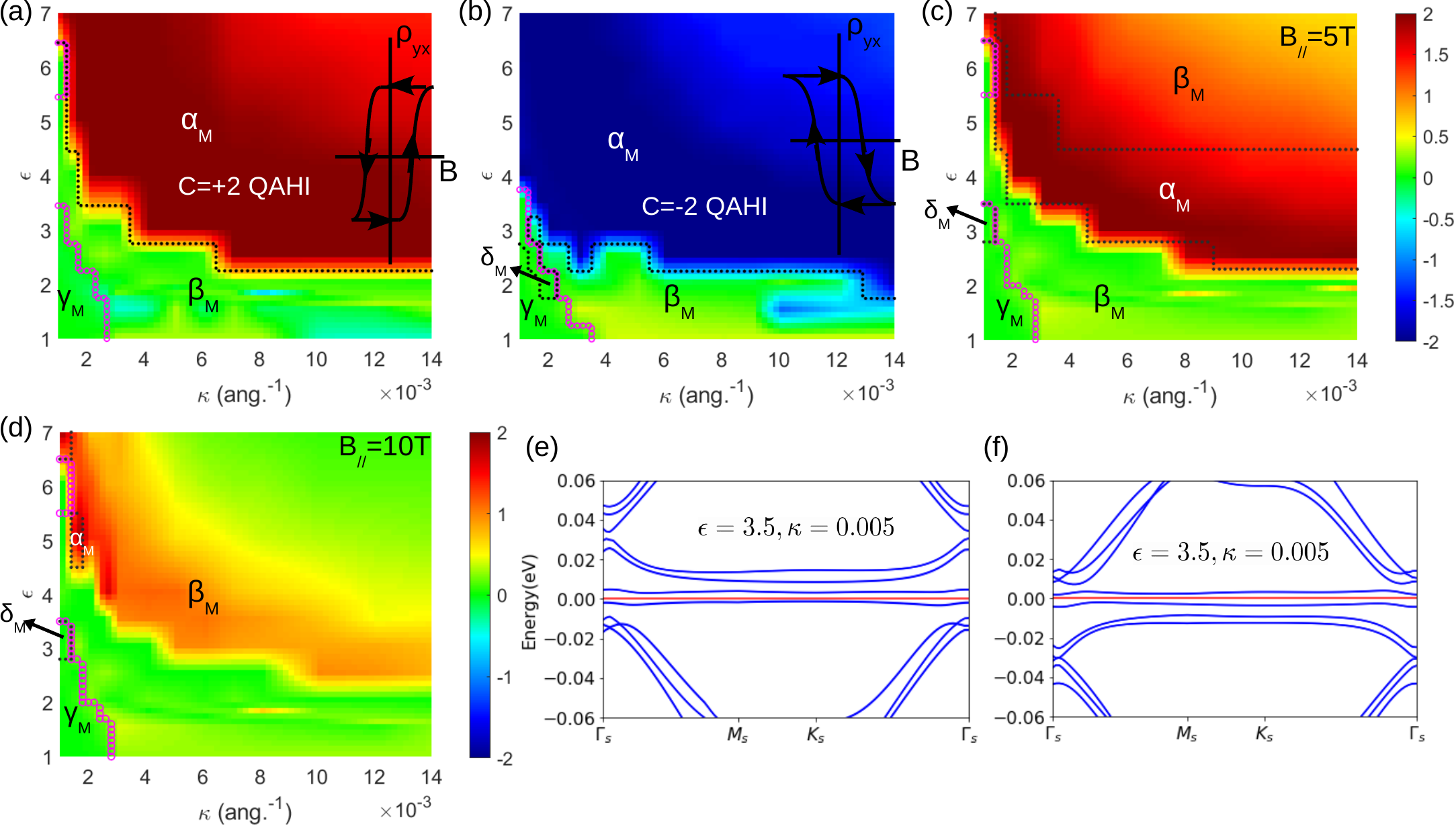}
\caption{The Hartree-Fock phase diagrams for hBN-aligned TBG at $\pm$ 1/2 fillings are shown in (a)-(d): (a), $-1/2$ filling, with in-plane magnetic field $B_{\parallel}\!=\!0$, (b) 1/2 filling,  $B_{\parallel}\!=\!0$, (c) -1/2 filling, $B_{\parallel}\!=\!5\,$T, (d) -1/2 filling, $B_{\parallel}\!=\!10\,$T. The color coding in (a)-(d) denotes the calculated anomalous Hall conductivities of the Hartree-Fock ground states with local exchange approximation. Hartree-Fock bandstructures of the QAH states at $\epsilon=3.5$, $\kappa=0.005\,\angstrom^{-1}$, (e) at -1/2 filling, and (f) at 1/2 filling.}
\label{fig:qah-inplane}
\end{figure*}

In Fig.~\ref{fig:ahc-varyangle}(a) we show how  the anomalous Hall conductivities of hBN-aligned TBG at $\pm 1/2$ fillings are dependent on the twist angle $\theta$. We see that for $\epsilon\!=\!4$, $\kappa\!=\!0.005\,\angstrom^{-1}$,  the $C\!=\!\pm 2$ QAH states appear only at the magic angle $\theta\!=\!1.05^{\circ}$. At $\theta\!=\!1.0^{\circ}$ the system  is in QAH insulating states with Chern number  $\mp 1$  at $\pm 1/2$ fillings with $\epsilon\!=\!4$ and $\kappa\!=\!0.005\,\angstrom^{-1}$. We leave the details of such  QAH states  with Chern number $\mp 1$ for future study.  At the other twist angles, the system remain metallic  with large anomalous Hall conductivities. Since the bandwidth is minimal at the magic angle, the valley polarization induced by Coulomb interactions would be strong enough to open a global gap only if the system is around the magic angle.  In Fig.~\ref{fig:ahc-varyangle}(b) we show the dependence the AHC on the twist angle at $\pm 3/4$ fillings with $\epsilon\!=\!2.5, \kappa\!=\!0.005\,\angstrom^{-1}$. Again, as the bandwidth is minimal at the magic angle, the $C\!=\!\pm 1$ QAH states occur only at the magic angle. The system becomes metallic with significant anomalous Hall conductivities at other twist angles.

%\section{Possible quantum anomalous Hall states at $\pm 1/2$ filling in hBN-aligned TBG}
\vspace{12pt}
\begin{center}
\textbf{\large \IV\ Possible quantum anomalous Hall states at $\pm 1/2$ filling in hBN-aligned TBG}
\end{center}

In this section we discuss the possible quantum anomalous Hall (QAH) effect at $\pm 1/2$ fillings in hBN-aligned TBG at the magic angle. As discussed in the main text, the Hartree-Fock ground states (with local-exchange approximation) at $\pm 1/2$ fillings are QAH states with Chern number $\pm 2$ in the $\alpha_M$ phase. Such states are orbital ferromagnetic and spin paramagnetic states. The Hartree-Fock phase diagrams at $-1/2$ and $1/2$ fillings in the parameter space of the dielectric constant $\epsilon$ and the inverse screening length $\kappa$ are shown in Fig.~\ref{fig:qah-inplane}(a) and (b) respectively (also shown in Fig.~6(c)-(d) in the main text). The bandstructures of the QAH states with $\epsilon=3.5$, $\kappa=0.005\,\angstrom^{-1}$ at $-1/2$ and $1/2$ fillings are shown in Fig.~\ref{fig:qah-inplane}(e) and (f) respectively.

Such  $C\!=\!\pm 2$ QAH states at $\mp 1/2$ fillings exhibit unconventional responses to external magnetic fields. First,  the valley-polarized $C\!=\!\pm2$ QAH states at $\mp1/2$ fillings are orbital ferromagnetic states associated with giant spontaneous orbital magnetizations $\sim\!5 \hbox{-} 12\,\mu_{\textrm{B}}$ per moir\'e supercell. Thus a weak vertical external magnetic field will stablize the QAH states due to the coupling to the orbital magnetization.  
On the other hands, the valley-polarized $C\!=\!\pm 2$ QAH states are also \textit{spin paramagnetic} states, which implies that the spin Zeeman effect would compete with the topological gap generated by the valley polarization. It follows that if one applies an in-plane magnetic field which only couples to spin magnetization, the $C\!=\!\pm 2$ QAH states at $\mp 1/2$ fillings will be strongly suppressed.  In Fig.~\ref{fig:qah-inplane} (c) we show  the AHC of the HF ground states at $-1/2$ filling in the presence of an in-plane magnetic field $B_{//}\!=\!5\,$T. Clearly the region of the QAH phase has been significantly reduced due to the presence of the weak in-plane magnetic field. In Fig.~\ref{fig:qah-inplane}(d) we show the calculated AHC of the HF ground states at $-1/2$ filling with the in-plane magnetic field $B_{//}\!=\!10\,$T. We see that the QAH phase has been completely ruled out from the phase diagram.  In the $C\!=\!\pm2$ QAH phase without magnetic field, the calculated Chern-insulating gap $\sim 5\hbox{-}10\,$meV; on the other hand, the spin Zeeman splitting with $B_{//}\!\lessapprox\!10\,$T is less than 1\,meV. This implies that the spin Zeeman splittings have been significantly enhanced by Coulomb interactions, such that the QAH phase can be destructed by in-plane magnetic fields by such weak magnetic fields.

\end{document}